%% file: KLM_BOW.tex
\documentclass[aps,prb,reprint,superscriptaddress,showpacs]{revtex4-1}
\usepackage{graphicx}
\usepackage{subfigure}
\usepackage{dcolumn}
\usepackage{bm}
\usepackage{hyperref}
\usepackage{indentfirst}
\usepackage{braket}
\usepackage{amsmath}
\usepackage{amssymb}
\hypersetup{colorlinks=true, citecolor=blue, urlcolor=blue, linkcolor=blue}
\begin{document}

\title{Coupled dimer and bond-order-wave order in the quarter-filled one-dimensional Kondo lattice model}


\author{Yixuan Huang}
\affiliation{Texas Center for Superconductivity, University of Houston, Houston, Texas 77204, USA.}
\affiliation{Department of Physics and Astronomy, California State University, Northridge, California 91330, USA.}
\author{D. N. Sheng}
\affiliation{Department of Physics and Astronomy, California State University, Northridge, California 91330, USA.}
\author{C. S. Ting}
\affiliation{Texas Center for Superconductivity, University of Houston, Houston, Texas 77204, USA.}


\date{\today}

\begin{abstract}
Motivated by the experiments on the organic compound $(Per)_{2}[Pt(mnt)_{2}]$, we study the ground state of the one-dimensional Kondo lattice model at quarter filling with the density matrix renormalization group method. We show a coupled dimer and bond-order-wave (BOW) state in the weak coupling regime for the localized spins and itinerant electrons, respectively. The quantum phase transitions for the dimer and the BOW orders occur at the same critical coupling parameter $J_{c}$, with the opening of a charge gap. The emergence of the combination of dimer and BOW order agrees with the experimental findings of the simultaneous Peierls and spin-Peierls transitions at low temperatures, which provides a theoretical understanding of such phase transition. We also show that the localized spins in this insulating state have quasi-long ranged spin correlations with collinear configurations, which resemble the classical dimer order in the absence of a magnetic order.
\end{abstract}


\maketitle

\section{Introduction}

Since the discover of the Krogmann salts~\cite{comes1973evidence}, many quasi-one-dimensional materials have been found to show various spin and charge orders~\cite{pouget2012bond} at low temperatures such as the blue bronzes $K_{0.3}MoO_{3}$~\cite{travaglini1984charge,pouget1991neutron}, the transition metals $NbSe_{3}$ and $TaS_{3}$~\cite{schlenker1996physics}, the $CuGeO_{3}$~\cite{hase1993observation,hase1993magnetic}, and the 2:1 $D_{2}X$ organic salt~\cite{pouget2017peierls}. Interestingly, the $(Per)_{2}[Pt(mnt)_{2}]$ that contains quarter-filled metallic chains and half-filled insulating chains shows a unique combination of charge and spin order at almost the same transition temperature~\cite{henriques1984electrical,henriques1986electrical,bourbonnais1991nuclear,matos1996modification,green2011interaction}, which indicates that the transition is driven by the coupling between the two chains.

This coupling effect could be best described by the one-dimensional Kondo lattice (KL) model~\cite{doniach1977kondo,lacroix1979phase,fazekas1991magnetic,tsunetsugu1993phase,tsunetsugu1997ground} that consists of itinerant electrons coupled with the periodic localized spins through the Kondo coupling parameter $J$. In the large $J$ regime the ground state phase diagram is dominant by the ferromagnetism~\cite{mcculloch2002localized,peters2012ferromagnetic} as a result of the Kondo effect. At small $J$ the Kondo effect is suppressed and the phase belongs to a universal class of Tomonaga-Luttinger liquid at generic fillings with zero charge and spin gap~\cite{shibata1996friedel,shibata1997one}. However, for certain $J$ at the commensurate fillings of $n=\frac{1}{2}$~\cite{xavier2003dimerization}, $n=\frac{3}{4}$~\cite{huang2019charge}, and $n=1$~\cite{tsunetsugu1992spin} the state becomes insulating, and the insulating state at the quarter filling ($n=\frac{1}{2}$) is mostly related to the experimental results of the $(Per)_{2}[Pt(mnt)_{2}]$. 

The one-dimensional KL model at $n=\frac{1}{2}$ has been theoretically investigated to suggest an insulating state with semiclassical collinear spin configurations that resemble the dimer order at extremely small J~\cite{tsvelik2019physics}. Whether such state can be stabilized taking full account of the quantum fluctuations remain an open question, because quantum fluctuations are especially enhanced in low dimensionality. In addition, numerical studies have shown an insulating state with dimerization of the localized spins that survives up to an intermediate J~\cite{xavier2003dimerization,PhysRevB.78.144406}. However, certain competing orders may appear in the itinerant electrons and the true nature of this state remains unexplored.

\begin{figure}
\centering
\includegraphics[width=0.8\columnwidth]{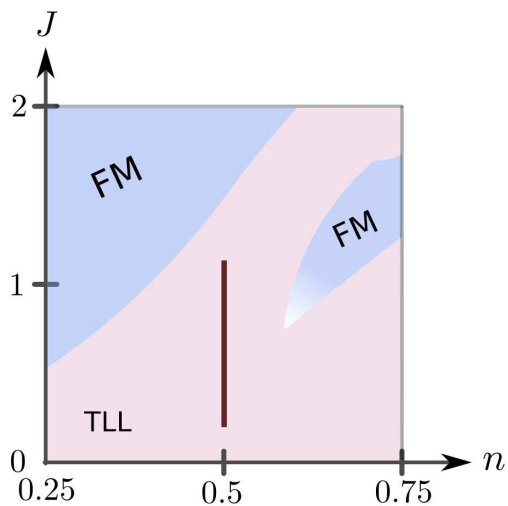}
\caption{\label{Fig:phase}(Color online) The schematic ground state phase diagram of the one-dimensional KL model for $0.25<n<0.75$ and $0<J<2$. The red regime at $n=\frac{1}{2}$ is the insulating dimer/BOW state; the blue one labeled FM is the ferromagnetic state; the pink one labeled TLL is the Tomonaga-Luttinger liquid.}
\end{figure}

Motivated by the experiments of quasi-one-dimensional organic compound $(Per)_{2}[Pt(mnt)_{2}]$, we study the ground state of the one-dimensional KL model with density matrix renormalization group (DMRG)~\cite{PhysRevLett.69.2863,PhysRevB.48.10345,schollwock2011density}. We provide detailed numerical evidence of the dimer order for the localized spins and discuss the nature of this dimer order. A coupled bond-order-wave (BOW) is also found in this regime with a simultaneous quantum phase transition along with the dimer order at the critical $J_{c}$, and we discuss the possible connections between our results and experimental observations.

We consider the standard KL Hamiltonian with isotropic coupling term, which is given as 

\begin{equation}
\label{eq:H}
\mathcal{H}=-t{\displaystyle {\displaystyle {\textstyle {\displaystyle \sum_{i=1,\sigma}^{L-1}c_{i,\sigma}^{\dagger}}c_{i+1,\sigma}+H.c.}}+J\sum_{i=1}^{L}\overrightarrow{S_{i}}\cdot\overrightarrow{s_{i}}}
\end{equation}

where $c_{i,\sigma}^{\dagger}$ is the electron creation operator on site $i$ with spin index $\sigma$, and is summed over the system length $L$; the $\overrightarrow{S_{i}}$ is the localized spin-$\frac{1}{2}$ operator; the $\overrightarrow{s_{i}}=\frac{1}{2}\sum_{\alpha,\beta}c_{i,\alpha}^{\dagger}\overrightarrow{\sigma}_{\alpha,\beta}c_{i,\beta}$ represents the conduction electron spin operator with $\overrightarrow{\sigma}$ being the Pauli matrices in the spin space. The hopping parameter $t$ and the lattice spacing $\xi$ is set to 1 for the rest of the paper unless noted otherwise. 

Our main results focus on the quarter filling with the intermediate coupling $J$ as illustrated in Fig.\ref{Fig:phase}. First, we obtain a finite order parameter for both dimer and BOW order after finite size extrapolation for various $J$ and identify a simultaneous transition point at $J^{Dimer}_{c}=J^{BOW}_{c}\approx 1.2$. In order to reduce the effect by the open boundary condition, we show the same order using both finite and infinite DMRG (iDMRG) methods in this regime (see Appendix \ref{iDMRG}). In addition, we show quasi-long ranged spin-spin correlations of the localized spins with patterns similar to the semiclassical prediction of $\upuparrows \downdownarrows$~\cite{tsvelik2019physics}, without breaking the spin $S(U)_{2}$ symmetry. Meanwhile the dimer correlation remains an exponential decay function, which is qualitatively different from the dimerized phase at the Majumdar-Ghosh point in the Heisenberg model\cite{haldane1982spontaneous}. Our results represent important progress in understanding the collective behavior of localized spins coupled to the itinerant electrons in one or quasi-one-dimensional systems.

We use finite size and infinite $U(1)$ DMRG method~\cite{ITensorandTenPy,tenpy} with 5000 - 8000 states kept in order to reach the truncation error at around $10^{-8}$. For finite DMRG various lengths are used for the finite size extrapolation with open boundary conditions, and the physical observables such as the spin correlations are extracted using the middle half of the chain in order to minimize the boundary effect. Various states kept and lengths are tested to ensure the numerical convergence of the results. 

The rest of the paper is organized as follows: In Sec.\ref{BOW}, we present the numerical evidence for the coupled BOW and dimer order in the intermediate $J$ at $n=\frac{1}{2}$, as well as the breakdown of such order in the presence of a large Zeeman field. We further confirm such phase with a finite charge gap in Sec.\ref{gap}. In Sec.\ref{correlation} we study the spin correlation functions. Sec.\ref{summary} contains discussions and summary.

\section{Coupled dimer and BOW order}\label{BOW}

\begin{figure}
\centering
\input{Dim_BOW_J_combined.tex}
\caption{\label{Fig:BOWDim}(Color online) The local dimer order (a) and BOW order (b) is obtained on the finite chain of $L=112$ at $n=\frac{1}{2},J=0.6$. Only half of the lattice is shown due to inversion symmetry. The inset of (a) and (b) is the same quantity, respectively, plotted near the center to show the same wave vector of $q=\pi $. The finite size extrapolation of the dimer order (c) and the BOW order (d) is shown for several $J$ at $n=\frac{1}{2}$, where a least-square fit to the second order of polynomials in 1/L is used. The dimer order and BOW order after the extrapolation is given in (e) and (f) as a function of $J$, respectively.}
\end{figure}
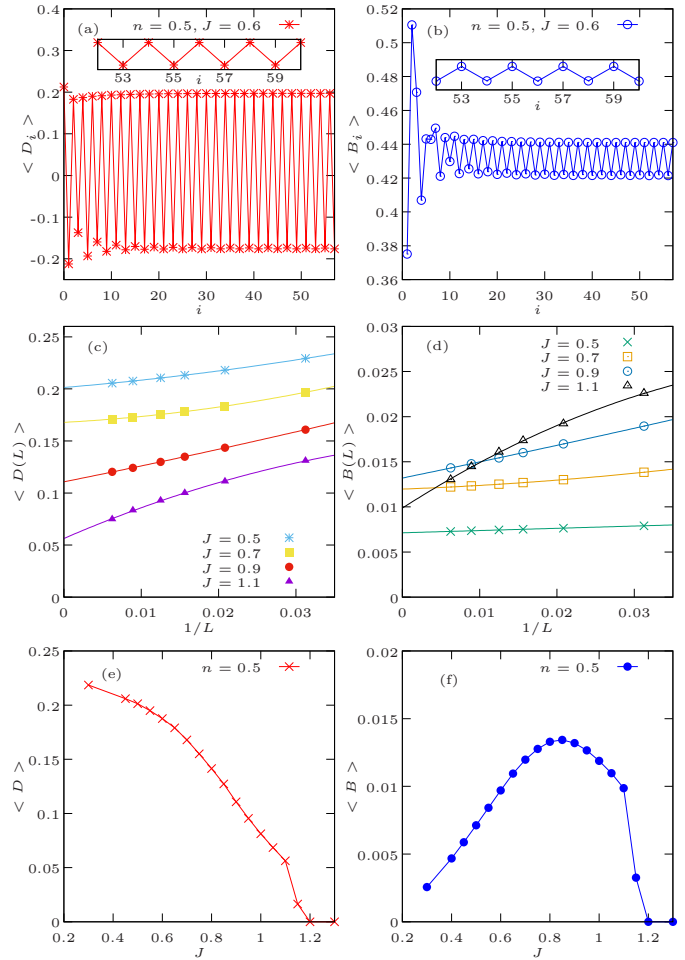

The simplest $spin-\frac{1}{2}$ dimer state consists of every neighboring pairs of spins forming an independent spin singlet state~\cite{white1996dimerization}. The order parameter of the dimer states $<D>$ is defined by the difference between neighboring spin bonds as given in Eq.\ref{eq:Dim_order}, where $\overrightarrow{S}_{i}$ refers to the localized spin at site $i$.

\begin{equation}
\label{eq:Dim_order}
\begin{split}
<D>& =\lim_{L\rightarrow \infty }<D(L)> \\
<D(L)>& =\frac{1}{L}\sum_{i}^{L}(-1)^{i}<D_{i}> \\
& =\frac{1}{L}\sum_{i}^{L}(-1)^{i} <\overrightarrow{S}_{i}\cdot \overrightarrow{S}_{i+1}>
\end{split}
\end{equation}

Following the original study of this dimer order in the one-dimensional KL model~\cite{xavier2003dimerization}, we show a robust dimer pattern in the real space without any pinning field in the intermediate coupling regime at $n=\frac{1}{2}$ for the localized spins. As an example given in Fig.\ref{Fig:BOWDim}(a), the dimer order becomes almost uniform away from the boundary. In respect to the concerns that this result may be an artificial effect caused by the open boundary, 
we also show the same order with iDMRG methods (see Appendix \ref{iDMRG}). Besides the dimer order, the BOW is identified in the same regime with the order parameter that is defined by the alternating electron hopping energy as below.

\begin{equation}
\label{eq:BOW_order}
\begin{split}
<B> & =\lim_{L\rightarrow \infty }<B(L)> \\
<B(L)> & =\frac{1}{L}\sum_{i}^{L}(-1)^{i}<B_{i}> \\
& =\frac{1}{2L}\sum_{i \sigma }^{L}(-1)^{i} <C^{\dagger }_{i\sigma }\cdot C_{i+1\sigma } + H.c.>
\end{split}
\end{equation}

As shown in Fig.\ref{Fig:BOWDim}(b) the alternating hopping energy between the neighboring sites also becomes uniform away from the boundary, suggesting an finite BOW order for the itinerant electrons. Meanwhile the electron density on every site remains the same away from the boundary.

The dimer and BOW orders are extrapolated into the thermodynamic limit using various lengths $L$ as shown in Fig.\ref{Fig:BOWDim}(c) and (d). The orders have a slight decay over the system length in the bulk of the dimer/BOW regime and remain finite after the extrapolation of $L\rightarrow \infty$. The extrapolated dimer order is given in Fig.\ref{Fig:BOWDim}(e), where it increases monotonically from 0 as $J$ decreases from $1.2$. The extrapolated BOW order also becomes finite at $J=1.2$ as shown in Fig.\ref{Fig:BOWDim}(f), and reaches maximum around $J=0.85$. For extremely small $J$ the numerical results are hard to converge, and we cannot be sure whether the order parameters goes to zero at finite $J$. However, based on the results we show a simultaneous quantum phase transition into the co-existing dimer and BOW state at $J_{c}\approx 1.2$. The two orders are coupled as they vanish at the same critical $J_{c}$.

In the presence of a strong external field $h$, the spins polarize and break the dimer order~\cite{xavier2003dimerization}. We have found that for $J=0.8$, the dimer order breaks down at around $h=0.05$, where the BOW order also vanishes. This further supports the co-existence of these two orders.

\section{Charge gap and spin gap}\label{gap}

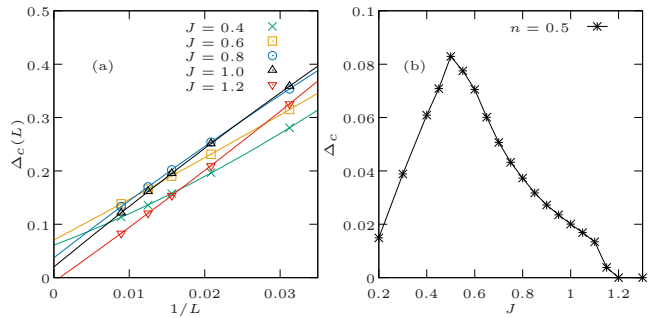
\begin{figure}
\centering
\input{Charge_gap_N_J.tex}
\caption{\label{Fig:Cgap}(Color online) (a) is the finite size extrapolation of the charge gap for various $J$ at $n=\frac{1}{2}$. We use a least-square fit to the second order of polynomials in 1/L. (b) is the extrapolated charge gap plotted against $J$.}
\end{figure}

Another quantity to separate this commensurate phase at $n=\frac{1}{2}$ from the phases at generic filling is the finite charge gap that is defined as 

\begin{equation}
\label{eq:Cgap}
\begin{split}
\Delta _{c}=lim_{L\rightarrow \infty } [ E_{0}\left ( N_{e}= N+2 \right )+E_{0} ( N_{e}=\\ 
N-2 )-2E_{0}\left ( N_{e}= N \right )  ]
\end{split}
\end{equation}
 
where $E_{0}\left ( N_{e} \right )$ is the ground state energy in the total electron number sector $N_{e}$. As shown in Fig.\ref{Fig:Cgap}(a), the charge gap is extrapolated into the thermodynamic limit in a similar way as the order parameters, and remains finite after the extrapolation in the dimer/BOW state. The extrapolated charge gap is plotted against $J$ in Fig.\ref{Fig:Cgap}(b), where it also becomes finite below $J_{c}=1.2$. The charge gap rises monotonically as $J$ decreases from 1.2 and reaches maximum at around $J=0.5$. The finite charge gap is consistent with the BOW order found in this regime.

The spin gap is obtained by the energy difference between the lowest states in $S=0$ and $1$ spin sectors, respectively. We calculate the spin gap in the co-existing dimer/BOW state with various system lengths. A very small spin gap (in the order of $10^{-4}$) is obtained after the finite size extrapolation, and it decreases with increasing states kept, indicating a vanishing spin gap. The localized spins have formed a dimer state, but the effective electron spin in one unit cell is $\frac{1}{2}$, which is consistent with the gapless spin excitation, resembling the physics of the $spin-\frac{1}{2}$ Heisenberg chain.

Furthermore, we study the entanglement entropy of the dimer/BOW state in the framework of the conformal field theory~\cite{calabrese2004entanglement}. As shown in Fig.\ref{Fig:VEECorrPeak}(a), the entanglement entropy is plotted against subsystem sizes at $n=\frac{1}{2},J=0.6$ for various lengths. The entanglement entropy shows a logarithmic dependence of the subsystem sizes that follows the general relation as below.

\begin{equation}
\label{eq:VEE}
S_{EE}(i)=\frac{c}{6}ln[\frac{L}{\pi }sin(\frac{i\pi }{L })]+g
\end{equation}

Here $S_{EE}(i)$ is the entanglement entropy of the subsystem with the length $i$. $L$ is the whole system length; $c$ is the central charge; $g$ is a non-universal constant. The fitting gives the central charge $c\approx 1$ for various system lengths in the dimer/BOW state, which is also consistent with the gapless spin excitations. 

\section{Spin-spin correlations}\label{correlation}

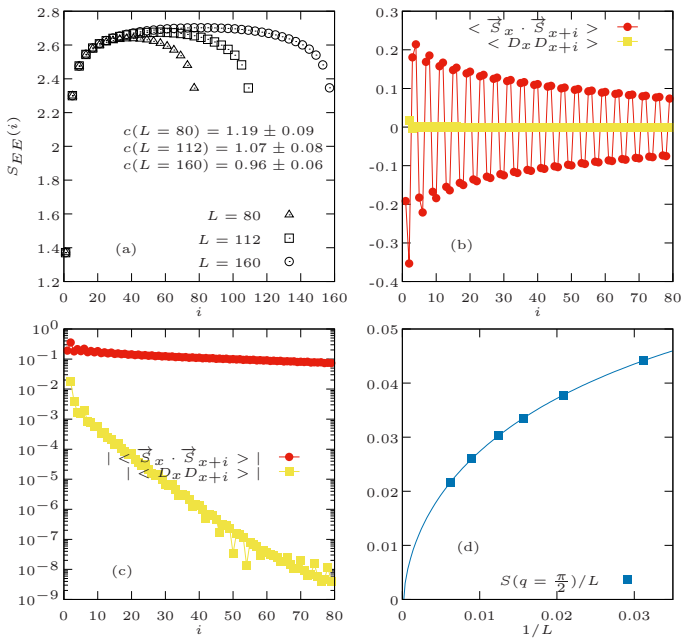
\begin{figure}
\centering
\input{VEE_Correlations_dimer_spin_peak_L.tex}
\caption{\label{Fig:VEECorrPeak}(Color online) (a) is the entanglement entropy obtained with various system lengths. The $S_{EE}(i)$ close to the left edge is fit by Eq.\ref{eq:VEE} to obtain the central charge $c$. The red line in (b) refers to the spin-spin correlations and the yellow line refers to the dimer-dimer correlations. The results of (b) are obtained on the chain of $L=160$ and $x$ is chosen to be $\frac{L}{4}$ in order to minimize the boundary effect. (c) is the same correlations with log scale in the y-axis showing the correlation decay over distance. (d) is the finite size extrapolation of the structure peak value. All results above are obtained at $n=\frac{1}{2}, J=0.6$.}
\end{figure}

The localized spins are mediated by the effective interactions known as the Ruderman-Kittel-Kasuya-Yosida (RKKY) interactions~\cite{ruderman1954indirect,kasuya1956theory,yosida1957magnetic}, which are long-ranged interactions with a staggered sign. Thus, the resulting dimer state is expected to be different from the dimerization at the Majumdar-Ghosh point in the $J_{1}-J_{2}$ Heisenberg model. To further reveal the nature of this dimer order, we obtain the spin-spin correlations of the localized spins, as shown in Fig.\ref{Fig:VEECorrPeak}(b) and (c). The spin-spin correlations in the dimer/BOW state decay slowly with the power-law behavior, and it oscillates with a period of 4 lattice constants, which is doubled by the period of the dimer order. This suggests that the dimer order originates from the quasi-long ranged correlations between the localized spins instead of the formation of a spin singlet state between neighboring sites. The collective behavior of the localized spins could be regarded as a quantum analogy of the classical dimer state (with spin configuration of $\upuparrows \downdownarrows$) without any magnetic order (see Appendix \ref{convergence}).

We further examine the nature of this dimer state by the dimer-dimer correlations, which are defined as the two point correlation functions of the neighboring spin bonds $ \left \langle D_{x}D_{x+i} \right \rangle $, with $D_{x}=\overrightarrow{S}_{x}\cdot \overrightarrow{S}_{x+1}$. As shown in Fig.\ref{Fig:VEECorrPeak}(b) and (c), the dimer-dimer correlation has an exponential decay over distance, and it is much smaller than the spin-spin correlation. This result is obtained under the open boundary condition on a finite size chain. A more complete study of the dimer-dimer correlation would be under the periodic boundary condition with vanishing dimer order~\cite{shibata2011boundary}, but we have found that it is hard to reach numerical convergence even for a small system length of $L=32$ under the periodic boundary condition. However, the robust dimer order in the absence of pinning field and the periodic pattern in spin-spin correlations are consistent with the dimer state in this regime.

To explore other possible spin order, we study the spin structure factor of the localized spins. The structure factor is defined as 

\begin{equation}
\label{eq:structure}
S\left ( q \right ) = \frac{1}{L}\sum_{i,j}\left \langle \overrightarrow{S}_{i}\cdot \overrightarrow{S}_{j} \right \rangle e^{i q \left ( r_{i} -  r_{j} \right ) }
\end{equation}

In the dimer/BOW state, the structure factor has only one dominant peak at $q=\frac{\pi }{2}$ which could be seen from Fig.\ref{Fig:VEECorrPeak}(b) with the period of 4 lattice constants. The peak value increases slowly with the system length $L$, and as shown in Fig.\ref{Fig:VEECorrPeak} (d), the peak value divided by the system length decreases rapidly and goes to zero in the thermodynamic limit, which is consistent with the absence of any magnetic order. This result is also expected from the general statement that a spontaneous breaking of $S(U)_{2}$ symmetry is forbidden for the one-dimensional system in the thermodynamic limit~\cite{mermin1966absence}.

\section{Discussions and Summary}\label{summary}

Several studies~\cite{xavier2003dimerization,PhysRevB.78.144406,tsvelik2019physics} have suggested the dimerization of localized spins at $n=\frac{1}{2}$ in the one-dimensional KL model with a finite charge gap. The dimerization enlarges the effective unit cell by 2, which may result in a band insulator similar to the case at $n=1$. However, we argue that the finite charge gap is related to the emergent BOW, which is promoted by the electron backscattering induced by the localized spins. Indeed, a similar spin configuration to our spin correlation result is realized with semiclassical analysis considering such backscattering terms~\cite{tsvelik2019physics}. In addition, the dimer order can be stabilized by the emergence of the BOW with matching wave vectors. Because the same spin coupling is energetically favored by the RKKY interaction~\cite{litvinov1998rkky,rusin2017calculation}, the neighboring sites with smaller hopping bond energy have more tendency to form parallel localized spins.

The emergence of this co-existing dimer and BOW state depends crucially on the commensurate electron filling, which may raise concerns about whether the state remains stable under a perturbation. In particular, previous numerical studies of this state are conducted on a finite size lattice with open boundary condition. With a large chemical potential applying on the edge the dimer order becomes weaker and in general a short ranged order may be induced by the open boundary such as the Friedel oscillations of the electron density~\cite{hotta2006absence,shibata2011boundary}. Here we have shown the dimer/BOW order with almost the same value on an infinitely long chain (see Appendix \ref{iDMRG}), indicating that the order is robust. Also, the DMRG algorithm only targets for the lowest state at a given electron filling because of the conservation of total electrons, while in the grand canonical ensemble this state may survive in a small range of electron fillings~\cite{giamarchi2003quantum}.

The organic compound $(Per)_{2}[Pt(mnt)_{2}]$ is shown to be the experimental realization of the one-dimensional KL model~\cite{pouget2017peierls}, and could be best explained by our results. In particular, the perylene chain in the compound is metallic with quarter-filled electrons and the $Pt(mnt)_{2}$ chain is insulating with half-filled electrons that could be considered as localized spins. The interactions between the electrons in these two different chains make it an effective one-dimensional weakly coupled KL at $n=\frac{1}{2}$. At low temperatures this compound shows a combination of the dimer phase for the insulating stack~\cite{henriques1984electrical,green2011interaction}, and the charge-density-wave (CDW) phase for the metallic stack~\cite{matos1996modification,graf2004suppression,graf2004high}. The experimental measurement of different properties shows that the dimer and CDW phase transition occurs at almost the same temperature~\cite{gama1993interplay,bonfait1993spin}. Our numerical results also show a simultaneous phase transition at intermediate $J_{c}$. The quantitative difference between the BOW and the CDW is just the position of the peak of the wave, thus may appear similarly in the experiments as electron density modulations; see review [\cite{pouget2012bond}]. In addition, the spin-Peierls transition into the dimer phase agrees with the finite dimer order identified in this regime. 

However, the dimer/BOW state identified here has a matching wave vector while the experimental results show that the wave vector of the perylene chain with CDW order and the Pt chain with dimer order differs by a factor of 2 ($q^{Pt}=2q^{per}$). This difference is not understood, which could be an effect of more chains weakly coupled together in experimental systems. Future studies may include two itinerant electron chains coupled by the magnetic impurities.

To summarize, we have numerically identified a coupled dimer and BOW order for the localized spins and itinerant electrons, respectively, in the one-dimensional KL model at $n=\frac{1}{2}$. The study of its evolution with the Kondo coupling $J$ shows a simultaneous quantum phase transition of both orders at $J_{c} \approx 1.2$. This result agrees quantitatively with the experimental findings on the organic compound $(Per)_{2}[Pt(mnt)_{2}]$. The localized spins are mediated by the RKKY interactions that results in a correlated dimer state with quasi-long ranged spin-spin correlations. This dimer state is qualitatively different from the dimer phase in the extended Heisenberg model where every neighboring spin pairs form a singlet state. Although the results are restricted in one dimension, it provides a new example of the interaction driven phase transition and a hindsight to the Kondo physics.

\begin{acknowledgments}
Y.H and C.S.T were supported by the Texas Center for Superconductivity and the Robert A. Welch Foundation Grant No. E-1146. Work at CSUN was supported by the U.S. Department of Energy, Office of Basic Energy Sciences under the grant No. DE-FG02-06ER46305 (Y.H, D.N.S). Numerical calculations were completed in part with resources provided by the Center for Advanced Computing and Data Science at the University of Houston. 
\end{acknowledgments}


\appendix
\section{Numerical convergence}\label{convergence}

\begin{figure}
\centering
\input{states_converge.tex}
\caption{\label{FigS:converge}(Color online) The ground state energy (a), the dimer order (b), the BOW order (c), and the localized spin value $<S^{z}_{i}>$ at the center (d) for various states kept in the dimer/BOW state.}
\end{figure}
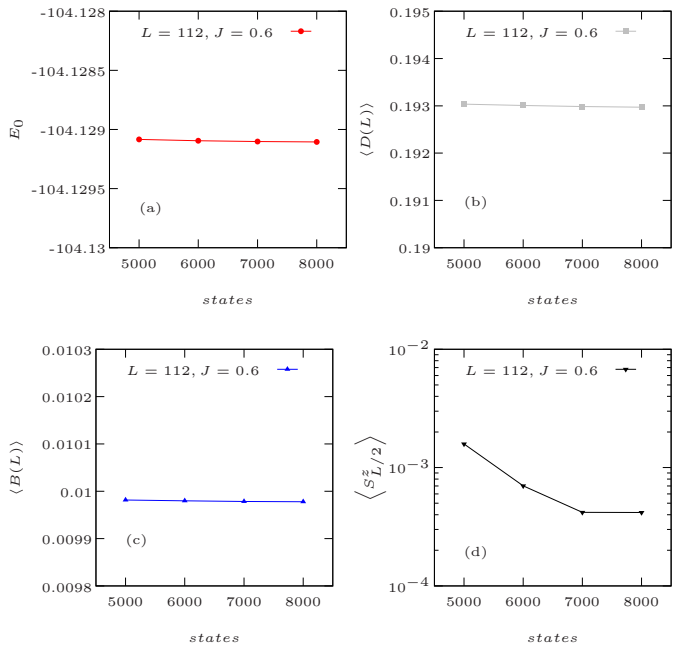

We ensure the convergence of the DMRG results by checking various quantities with increasing states kept. As an example at $J=0.6$ shown in Fig.\ref{FigS:converge}, the ground state energy remains almost unchanged over increasing states, as well as the dimer and BOW order. We have also found a very small localized spin value $<S^{z}_{i}>$ that decreases with increasing states kept, indicating the absence of a magnetic order.

For most of our calculations in the insulating dimer/BOW state, 5000 states are used in order to achieve a truncation error of $10^{-8}$. Generally, the convergence is harder to reach for smaller J, thus more states are needed.

\section{infinite DMRG results}\label{iDMRG}

\begin{figure}
\centering
\input{iDMRG_energy_j08.tex}
\caption{\label{FigS:iDMRG}(Color online) The ground state energy per site with increasing states kept, obtained with finite size DMRG and iDMRG at $n=\frac{1}{2}, J=0.8$. For the finite size DMRG method the energy per site is obtained through the center half of the chain to minimize the boundary effect.}
\end{figure}
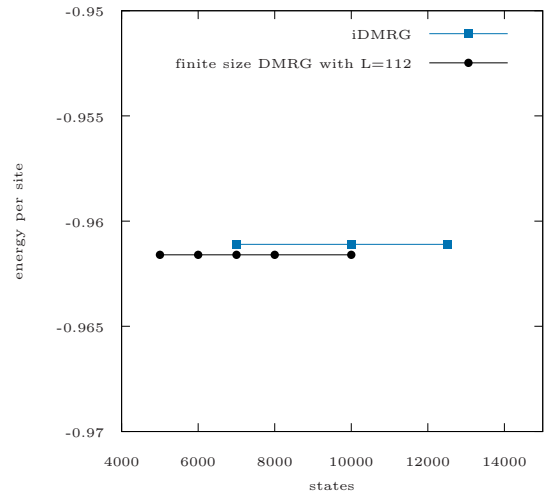

We show the same order parameters in the dimer/BOW state with iDMRG methods. The ground state energy per site is obtained by both finite DMRG and iDMRG methods with increasing states at the same parameters. As given in Fig.\ref{FigS:iDMRG}, the energy obtained by the two methods are very close (the slight difference is due to finite size effect) and remains almost the same with increasing states, indicating the convergence of the results. Under the iDMRG methods, the ground state at $n=\frac{1}{2}, J=0.8$ has a uniform alternating localized spin bonds with $<D>=0.1427$ and electron hopping energy with $<B>=0.0134$, which only differs by 1\% from the finite size DMRG results.

\bibliography{KLM_BOW}

\end{document}

%% file: Dim_BOW_J_combined.tex
\begingroup
  \makeatletter
  \providecommand\color[2][]{%
    \GenericError{(gnuplot) \space\space\space\@spaces}{%
      Package color not loaded in conjunction with
      terminal option `colourtext'%
    }{See the gnuplot documentation for explanation.%
    }{Either use 'blacktext' in gnuplot or load the package
      color.sty in LaTeX.}%
    \renewcommand\color[2][]{}%
  }%
  \providecommand\includegraphics[2][]{%
    \GenericError{(gnuplot) \space\space\space\@spaces}{%
      Package graphicx or graphics not loaded%
    }{See the gnuplot documentation for explanation.%
    }{The gnuplot epslatex terminal needs graphicx.sty or graphics.sty.}%
    \renewcommand\includegraphics[2][]{}%
  }%
  \providecommand\rotatebox[2]{#2}%
  \@ifundefined{ifGPcolor}{%
    \newif\ifGPcolor
    \GPcolortrue
  }{}%
  \@ifundefined{ifGPblacktext}{%
    \newif\ifGPblacktext
    \GPblacktexttrue
  }{}%
  \let\gplgaddtomacro\g@addto@macro
  \gdef\gplbacktext{}%
  \gdef\gplfronttext{}%
  \makeatother
  \ifGPblacktext
    \def\colorrgb#1{}%
    \def\colorgray#1{}%
  \else
    \ifGPcolor
      \def\colorrgb#1{\color[rgb]{#1}}%
      \def\colorgray#1{\color[gray]{#1}}%
      \expandafter\def\csname LTw\endcsname{\color{white}}%
      \expandafter\def\csname LTb\endcsname{\color{black}}%
      \expandafter\def\csname LTa\endcsname{\color{black}}%
      \expandafter\def\csname LT0\endcsname{\color[rgb]{1,0,0}}%
      \expandafter\def\csname LT1\endcsname{\color[rgb]{0,1,0}}%
      \expandafter\def\csname LT2\endcsname{\color[rgb]{0,0,1}}%
      \expandafter\def\csname LT3\endcsname{\color[rgb]{1,0,1}}%
      \expandafter\def\csname LT4\endcsname{\color[rgb]{0,1,1}}%
      \expandafter\def\csname LT5\endcsname{\color[rgb]{1,1,0}}%
      \expandafter\def\csname LT6\endcsname{\color[rgb]{0,0,0}}%
      \expandafter\def\csname LT7\endcsname{\color[rgb]{1,0.3,0}}%
      \expandafter\def\csname LT8\endcsname{\color[rgb]{0.5,0.5,0.5}}%
    \else
      \def\colorrgb#1{\color{black}}%
      \def\colorgray#1{\color[gray]{#1}}%
      \expandafter\def\csname LTw\endcsname{\color{white}}%
      \expandafter\def\csname LTb\endcsname{\color{black}}%
      \expandafter\def\csname LTa\endcsname{\color{black}}%
      \expandafter\def\csname LT0\endcsname{\color{black}}%
      \expandafter\def\csname LT1\endcsname{\color{black}}%
      \expandafter\def\csname LT2\endcsname{\color{black}}%
      \expandafter\def\csname LT3\endcsname{\color{black}}%
      \expandafter\def\csname LT4\endcsname{\color{black}}%
      \expandafter\def\csname LT5\endcsname{\color{black}}%
      \expandafter\def\csname LT6\endcsname{\color{black}}%
      \expandafter\def\csname LT7\endcsname{\color{black}}%
      \expandafter\def\csname LT8\endcsname{\color{black}}%
    \fi
  \fi
    \setlength{\unitlength}{0.0500bp}%
    \ifx\gptboxheight\undefined%
      \newlength{\gptboxheight}%
      \newlength{\gptboxwidth}%
      \newsavebox{\gptboxtext}%
    \fi%
    \setlength{\fboxrule}{0.5pt}%
    \setlength{\fboxsep}{1pt}%
\begin{picture}(5102.00,7652.00)%
    \gplgaddtomacro\gplbacktext{%
      \csname LTb\endcsname
      \put(279,5383){\makebox(0,0)[r]{\strut{}\tiny -0.2}}%
      \put(279,5697){\makebox(0,0)[r]{\strut{}\tiny -0.1}}%
      \put(279,6011){\makebox(0,0)[r]{\strut{}\tiny 0}}%
      \put(279,6326){\makebox(0,0)[r]{\strut{}\tiny 0.1}}%
      \put(279,6640){\makebox(0,0)[r]{\strut{}\tiny 0.2}}%
      \put(279,6954){\makebox(0,0)[r]{\strut{}\tiny 0.3}}%
      \put(279,7268){\makebox(0,0)[r]{\strut{}\tiny 0.4}}%
      \put(306,5116){\makebox(0,0){\strut{}\tiny 0}}%
      \put(664,5116){\makebox(0,0){\strut{}\tiny 10}}%
      \put(1022,5116){\makebox(0,0){\strut{}\tiny 20}}%
      \put(1380,5116){\makebox(0,0){\strut{}\tiny 30}}%
      \put(1738,5116){\makebox(0,0){\strut{}\tiny 40}}%
      \put(2095,5116){\makebox(0,0){\strut{}\tiny 50}}%
      \put(403,7127){\makebox(0,0)[l]{\strut{}\tiny (a)}}%
    }%
    \gplgaddtomacro\gplfronttext{%
      \csname LTb\endcsname
      \put(20,6247){\rotatebox{-270}{\makebox(0,0){\strut{}\tiny $ <D_{i}>$}}}%
      \put(1326,5006){\makebox(0,0){\strut{}\tiny $i$}}%
      \csname LTb\endcsname
      \put(1795,7150){\makebox(0,0)[r]{\strut{}\tiny $n=0.5, J=0.6$}}%
    }%
    \gplgaddtomacro\gplbacktext{%
      \csname LTb\endcsname
      \put(752,6755){\makebox(0,0){\strut{}\tiny 53}}%
      \put(1135,6755){\makebox(0,0){\strut{}\tiny 55}}%
      \put(1517,6755){\makebox(0,0){\strut{}\tiny 57}}%
      \put(1900,6755){\makebox(0,0){\strut{}\tiny 59}}%
    }%
    \gplgaddtomacro\gplfronttext{%
      \csname LTb\endcsname
      \put(1326,6744){\makebox(0,0){\strut{}\tiny $i$}}%
    }%
    \gplgaddtomacro\gplbacktext{%
      \csname LTb\endcsname
      \put(2830,5226){\makebox(0,0)[r]{\strut{}\tiny 0.36}}%
      \put(2830,5481){\makebox(0,0)[r]{\strut{}\tiny 0.38}}%
      \put(2830,5737){\makebox(0,0)[r]{\strut{}\tiny 0.4}}%
      \put(2830,5992){\makebox(0,0)[r]{\strut{}\tiny 0.42}}%
      \put(2830,6247){\makebox(0,0)[r]{\strut{}\tiny 0.44}}%
      \put(2830,6502){\makebox(0,0)[r]{\strut{}\tiny 0.46}}%
      \put(2830,6758){\makebox(0,0)[r]{\strut{}\tiny 0.48}}%
      \put(2830,7013){\makebox(0,0)[r]{\strut{}\tiny 0.5}}%
      \put(2830,7268){\makebox(0,0)[r]{\strut{}\tiny 0.52}}%
      \put(2857,5116){\makebox(0,0){\strut{}\tiny 0}}%
      \put(3215,5116){\makebox(0,0){\strut{}\tiny 10}}%
      \put(3572,5116){\makebox(0,0){\strut{}\tiny 20}}%
      \put(3930,5116){\makebox(0,0){\strut{}\tiny 30}}%
      \put(4288,5116){\makebox(0,0){\strut{}\tiny 40}}%
      \put(4646,5116){\makebox(0,0){\strut{}\tiny 50}}%
      \put(3036,7115){\makebox(0,0)[l]{\strut{}\tiny (b)}}%
    }%
    \gplgaddtomacro\gplfronttext{%
      \csname LTb\endcsname
      \put(2478,6247){\rotatebox{-270}{\makebox(0,0){\strut{}\tiny $ <B_{i}>$}}}%
      \put(3876,5006){\makebox(0,0){\strut{}\tiny $i$}}%
      \csname LTb\endcsname
      \put(4345,7150){\makebox(0,0)[r]{\strut{}\tiny $n=0.5, J=0.6$}}%
    }%
    \gplgaddtomacro\gplbacktext{%
      \csname LTb\endcsname
      \put(3303,6602){\makebox(0,0){\strut{}\tiny 53}}%
      \put(3685,6602){\makebox(0,0){\strut{}\tiny 55}}%
      \put(4068,6602){\makebox(0,0){\strut{}\tiny 57}}%
      \put(4450,6602){\makebox(0,0){\strut{}\tiny 59}}%
    }%
    \gplgaddtomacro\gplfronttext{%
      \csname LTb\endcsname
      \put(3876,6547){\makebox(0,0){\strut{}\tiny $i$}}%
    }%
    \gplgaddtomacro\gplbacktext{%
      \csname LTb\endcsname
      \put(279,2831){\makebox(0,0)[r]{\strut{}\tiny 0}}%
      \put(279,3224){\makebox(0,0)[r]{\strut{}\tiny 0.05}}%
      \put(279,3616){\makebox(0,0)[r]{\strut{}\tiny 0.1}}%
      \put(279,4009){\makebox(0,0)[r]{\strut{}\tiny 0.15}}%
      \put(279,4402){\makebox(0,0)[r]{\strut{}\tiny 0.2}}%
      \put(279,4794){\makebox(0,0)[r]{\strut{}\tiny 0.25}}%
      \put(306,2721){\makebox(0,0){\strut{}\tiny 0}}%
      \put(889,2721){\makebox(0,0){\strut{}\tiny 0.01}}%
      \put(1472,2721){\makebox(0,0){\strut{}\tiny 0.02}}%
      \put(2055,2721){\makebox(0,0){\strut{}\tiny 0.03}}%
      \put(481,4732){\makebox(0,0)[l]{\strut{}\tiny (c)}}%
    }%
    \gplgaddtomacro\gplfronttext{%
      \csname LTb\endcsname
      \put(-46,3852){\rotatebox{-270}{\makebox(0,0){\strut{}\tiny $ <D(L)>$}}}%
      \put(1326,2611){\makebox(0,0){\strut{}\tiny $1/L$}}%
      \csname LTb\endcsname
      \put(1795,3279){\makebox(0,0)[r]{\strut{}\tiny $J=0.5$}}%
      \csname LTb\endcsname
      \put(1795,3169){\makebox(0,0)[r]{\strut{}\tiny $J=0.7$}}%
      \csname LTb\endcsname
      \put(1795,3059){\makebox(0,0)[r]{\strut{}\tiny $J=0.9$}}%
      \csname LTb\endcsname
      \put(1795,2949){\makebox(0,0)[r]{\strut{}\tiny $J=1.1$}}%
    }%
    \gplgaddtomacro\gplbacktext{%
      \csname LTb\endcsname
      \put(2830,2831){\makebox(0,0)[r]{\strut{}\tiny 0}}%
      \put(2830,3171){\makebox(0,0)[r]{\strut{}\tiny 0.005}}%
      \put(2830,3512){\makebox(0,0)[r]{\strut{}\tiny 0.01}}%
      \put(2830,3852){\makebox(0,0)[r]{\strut{}\tiny 0.015}}%
      \put(2830,4192){\makebox(0,0)[r]{\strut{}\tiny 0.02}}%
      \put(2830,4533){\makebox(0,0)[r]{\strut{}\tiny 0.025}}%
      \put(2830,4873){\makebox(0,0)[r]{\strut{}\tiny 0.03}}%
      \put(2857,2721){\makebox(0,0){\strut{}\tiny 0}}%
      \put(3440,2721){\makebox(0,0){\strut{}\tiny 0.01}}%
      \put(4022,2721){\makebox(0,0){\strut{}\tiny 0.02}}%
      \put(4605,2721){\makebox(0,0){\strut{}\tiny 0.03}}%
      \put(3032,4737){\makebox(0,0)[l]{\strut{}\tiny (d)}}%
    }%
    \gplgaddtomacro\gplfronttext{%
      \csname LTb\endcsname
      \put(2439,3852){\rotatebox{-270}{\makebox(0,0){\strut{}\tiny $ <B(L)>$}}}%
      \put(3876,2611){\makebox(0,0){\strut{}\tiny $1/L$}}%
      \csname LTb\endcsname
      \put(4345,4755){\makebox(0,0)[r]{\strut{}\tiny $J=0.5$}}%
      \csname LTb\endcsname
      \put(4345,4645){\makebox(0,0)[r]{\strut{}\tiny $J=0.7$}}%
      \csname LTb\endcsname
      \put(4345,4535){\makebox(0,0)[r]{\strut{}\tiny $J=0.9$}}%
      \csname LTb\endcsname
      \put(4345,4425){\makebox(0,0)[r]{\strut{}\tiny $J=1.1$}}%
    }%
    \gplgaddtomacro\gplbacktext{%
      \csname LTb\endcsname
      \put(279,382){\makebox(0,0)[r]{\strut{}\tiny 0}}%
      \put(279,791){\makebox(0,0)[r]{\strut{}\tiny 0.05}}%
      \put(279,1199){\makebox(0,0)[r]{\strut{}\tiny 0.1}}%
      \put(279,1608){\makebox(0,0)[r]{\strut{}\tiny 0.15}}%
      \put(279,2016){\makebox(0,0)[r]{\strut{}\tiny 0.2}}%
      \put(279,2425){\makebox(0,0)[r]{\strut{}\tiny 0.25}}%
      \put(306,272){\makebox(0,0){\strut{}\tiny 0.2}}%
      \put(677,272){\makebox(0,0){\strut{}\tiny 0.4}}%
      \put(1048,272){\makebox(0,0){\strut{}\tiny 0.6}}%
      \put(1419,272){\makebox(0,0){\strut{}\tiny 0.8}}%
      \put(1790,272){\makebox(0,0){\strut{}\tiny 1}}%
      \put(2161,272){\makebox(0,0){\strut{}\tiny 1.2}}%
      \put(584,2262){\makebox(0,0)[l]{\strut{}\tiny (e)}}%
    }%
    \gplgaddtomacro\gplfronttext{%
      \csname LTb\endcsname
      \put(-46,1403){\rotatebox{-270}{\makebox(0,0){\strut{}\tiny $ <D>$}}}%
      \put(1326,162){\makebox(0,0){\strut{}\tiny $J$}}%
      \csname LTb\endcsname
      \put(1795,2307){\makebox(0,0)[r]{\strut{}\tiny $n=0.5$}}%
    }%
    \gplgaddtomacro\gplbacktext{%
      \csname LTb\endcsname
      \put(2830,382){\makebox(0,0)[r]{\strut{}\tiny 0}}%
      \put(2830,893){\makebox(0,0)[r]{\strut{}\tiny 0.005}}%
      \put(2830,1404){\makebox(0,0)[r]{\strut{}\tiny 0.01}}%
      \put(2830,1914){\makebox(0,0)[r]{\strut{}\tiny 0.015}}%
      \put(2830,2425){\makebox(0,0)[r]{\strut{}\tiny 0.02}}%
      \put(2857,272){\makebox(0,0){\strut{}\tiny 0.2}}%
      \put(3228,272){\makebox(0,0){\strut{}\tiny 0.4}}%
      \put(3598,272){\makebox(0,0){\strut{}\tiny 0.6}}%
      \put(3969,272){\makebox(0,0){\strut{}\tiny 0.8}}%
      \put(4340,272){\makebox(0,0){\strut{}\tiny 1}}%
      \put(4711,272){\makebox(0,0){\strut{}\tiny 1.2}}%
      \put(3135,2221){\makebox(0,0)[l]{\strut{}\tiny (f)}}%
    }%
    \gplgaddtomacro\gplfronttext{%
      \csname LTb\endcsname
      \put(2465,1403){\rotatebox{-270}{\makebox(0,0){\strut{}\tiny $ <B>$}}}%
      \put(3876,162){\makebox(0,0){\strut{}\tiny $J$}}%
      \csname LTb\endcsname
      \put(4345,2307){\makebox(0,0)[r]{\strut{}\tiny $n=0.5$}}%
    }%
    \gplbacktext
    \put(0,0){\includegraphics{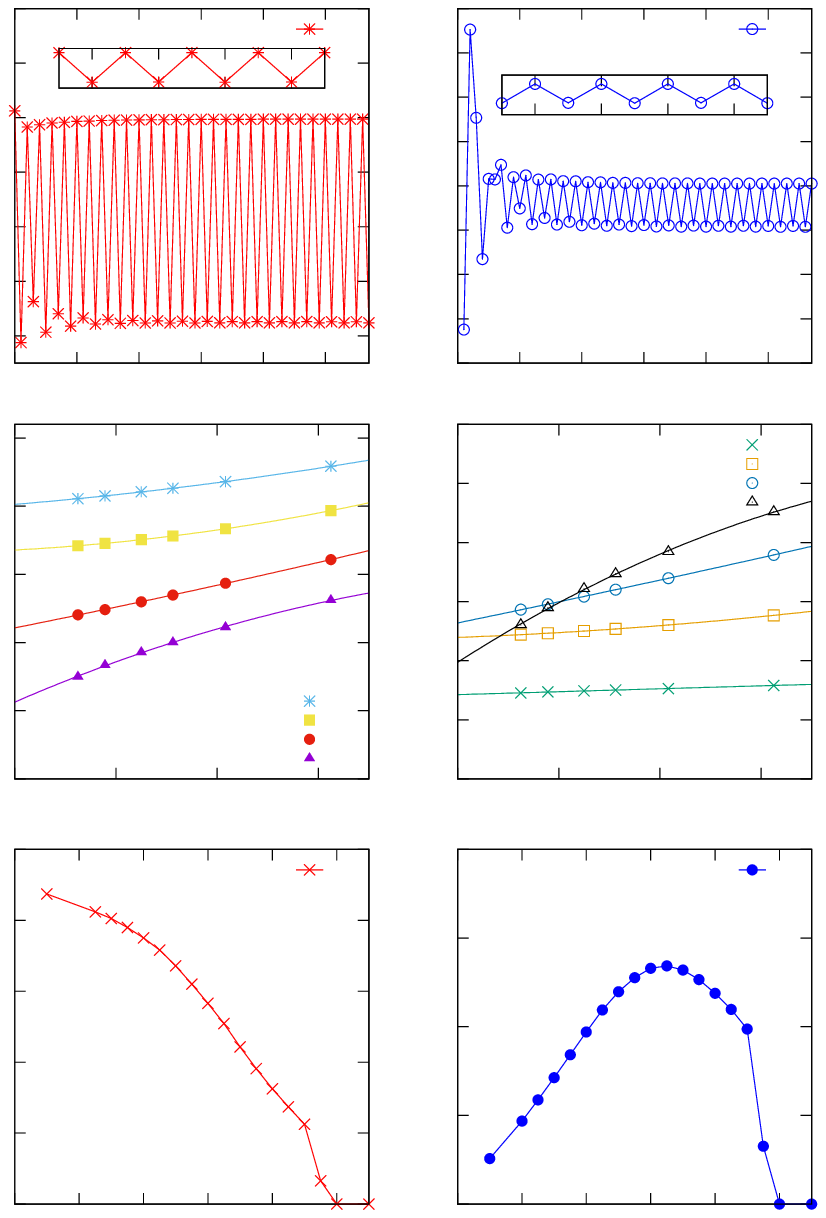}}%
    \gplfronttext
  \end{picture}%
\endgroup

%% file: Charge_gap_N_J.tex
\begingroup
  \makeatletter
  \providecommand\color[2][]{%
    \GenericError{(gnuplot) \space\space\space\@spaces}{%
      Package color not loaded in conjunction with
      terminal option `colourtext'%
    }{See the gnuplot documentation for explanation.%
    }{Either use 'blacktext' in gnuplot or load the package
      color.sty in LaTeX.}%
    \renewcommand\color[2][]{}%
  }%
  \providecommand\includegraphics[2][]{%
    \GenericError{(gnuplot) \space\space\space\@spaces}{%
      Package graphicx or graphics not loaded%
    }{See the gnuplot documentation for explanation.%
    }{The gnuplot epslatex terminal needs graphicx.sty or graphics.sty.}%
    \renewcommand\includegraphics[2][]{}%
  }%
  \providecommand\rotatebox[2]{#2}%
  \@ifundefined{ifGPcolor}{%
    \newif\ifGPcolor
    \GPcolortrue
  }{}%
  \@ifundefined{ifGPblacktext}{%
    \newif\ifGPblacktext
    \GPblacktexttrue
  }{}%
  \let\gplgaddtomacro\g@addto@macro
  \gdef\gplbacktext{}%
  \gdef\gplfronttext{}%
  \makeatother
  \ifGPblacktext
    \def\colorrgb#1{}%
    \def\colorgray#1{}%
  \else
    \ifGPcolor
      \def\colorrgb#1{\color[rgb]{#1}}%
      \def\colorgray#1{\color[gray]{#1}}%
      \expandafter\def\csname LTw\endcsname{\color{white}}%
      \expandafter\def\csname LTb\endcsname{\color{black}}%
      \expandafter\def\csname LTa\endcsname{\color{black}}%
      \expandafter\def\csname LT0\endcsname{\color[rgb]{1,0,0}}%
      \expandafter\def\csname LT1\endcsname{\color[rgb]{0,1,0}}%
      \expandafter\def\csname LT2\endcsname{\color[rgb]{0,0,1}}%
      \expandafter\def\csname LT3\endcsname{\color[rgb]{1,0,1}}%
      \expandafter\def\csname LT4\endcsname{\color[rgb]{0,1,1}}%
      \expandafter\def\csname LT5\endcsname{\color[rgb]{1,1,0}}%
      \expandafter\def\csname LT6\endcsname{\color[rgb]{0,0,0}}%
      \expandafter\def\csname LT7\endcsname{\color[rgb]{1,0.3,0}}%
      \expandafter\def\csname LT8\endcsname{\color[rgb]{0.5,0.5,0.5}}%
    \else
      \def\colorrgb#1{\color{black}}%
      \def\colorgray#1{\color[gray]{#1}}%
      \expandafter\def\csname LTw\endcsname{\color{white}}%
      \expandafter\def\csname LTb\endcsname{\color{black}}%
      \expandafter\def\csname LTa\endcsname{\color{black}}%
      \expandafter\def\csname LT0\endcsname{\color{black}}%
      \expandafter\def\csname LT1\endcsname{\color{black}}%
      \expandafter\def\csname LT2\endcsname{\color{black}}%
      \expandafter\def\csname LT3\endcsname{\color{black}}%
      \expandafter\def\csname LT4\endcsname{\color{black}}%
      \expandafter\def\csname LT5\endcsname{\color{black}}%
      \expandafter\def\csname LT6\endcsname{\color{black}}%
      \expandafter\def\csname LT7\endcsname{\color{black}}%
      \expandafter\def\csname LT8\endcsname{\color{black}}%
    \fi
  \fi
    \setlength{\unitlength}{0.0500bp}%
    \ifx\gptboxheight\undefined%
      \newlength{\gptboxheight}%
      \newlength{\gptboxwidth}%
      \newsavebox{\gptboxtext}%
    \fi%
    \setlength{\fboxrule}{0.5pt}%
    \setlength{\fboxsep}{1pt}%
\begin{picture}(5102.00,2550.00)%
    \gplgaddtomacro\gplbacktext{%
      \csname LTb\endcsname
      \put(279,357){\makebox(0,0)[r]{\strut{}\tiny 0}}%
      \put(279,760){\makebox(0,0)[r]{\strut{}\tiny 0.1}}%
      \put(279,1162){\makebox(0,0)[r]{\strut{}\tiny 0.2}}%
      \put(279,1565){\makebox(0,0)[r]{\strut{}\tiny 0.3}}%
      \put(279,1967){\makebox(0,0)[r]{\strut{}\tiny 0.4}}%
      \put(279,2370){\makebox(0,0)[r]{\strut{}\tiny 0.5}}%
      \put(306,247){\makebox(0,0){\strut{}\tiny 0}}%
      \put(874,247){\makebox(0,0){\strut{}\tiny 0.01}}%
      \put(1443,247){\makebox(0,0){\strut{}\tiny 0.02}}%
      \put(2011,247){\makebox(0,0){\strut{}\tiny 0.03}}%
      \put(590,1967){\makebox(0,0)[l]{\strut{}\tiny (a)}}%
    }%
    \gplgaddtomacro\gplfronttext{%
      \csname LTb\endcsname
      \put(20,1363){\rotatebox{-270}{\makebox(0,0){\strut{}\tiny $ \Delta _{c}(L)$}}}%
      \put(1300,137){\makebox(0,0){\strut{}\tiny $1/L$}}%
      \csname LTb\endcsname
      \put(1744,2252){\makebox(0,0)[r]{\strut{}\tiny $J=0.4$}}%
      \csname LTb\endcsname
      \put(1744,2142){\makebox(0,0)[r]{\strut{}\tiny $J=0.6$}}%
      \csname LTb\endcsname
      \put(1744,2032){\makebox(0,0)[r]{\strut{}\tiny $J=0.8$}}%
      \csname LTb\endcsname
      \put(1744,1922){\makebox(0,0)[r]{\strut{}\tiny $J=1.0$}}%
      \csname LTb\endcsname
      \put(1744,1812){\makebox(0,0)[r]{\strut{}\tiny $J=1.2$}}%
    }%
    \gplgaddtomacro\gplbacktext{%
      \csname LTb\endcsname
      \put(2728,357){\makebox(0,0)[r]{\strut{}\tiny 0}}%
      \put(2728,760){\makebox(0,0)[r]{\strut{}\tiny 0.02}}%
      \put(2728,1162){\makebox(0,0)[r]{\strut{}\tiny 0.04}}%
      \put(2728,1565){\makebox(0,0)[r]{\strut{}\tiny 0.06}}%
      \put(2728,1967){\makebox(0,0)[r]{\strut{}\tiny 0.08}}%
      \put(2728,2370){\makebox(0,0)[r]{\strut{}\tiny 0.1}}%
      \put(2755,247){\makebox(0,0){\strut{}\tiny 0.2}}%
      \put(3116,247){\makebox(0,0){\strut{}\tiny 0.4}}%
      \put(3478,247){\makebox(0,0){\strut{}\tiny 0.6}}%
      \put(3839,247){\makebox(0,0){\strut{}\tiny 0.8}}%
      \put(4201,247){\makebox(0,0){\strut{}\tiny 1}}%
      \put(4562,247){\makebox(0,0){\strut{}\tiny 1.2}}%
      \put(2936,1967){\makebox(0,0)[l]{\strut{}\tiny (b)}}%
    }%
    \gplgaddtomacro\gplfronttext{%
      \csname LTb\endcsname
      \put(2376,1363){\rotatebox{-270}{\makebox(0,0){\strut{}\tiny $ \Delta _{c}$}}}%
      \put(3749,137){\makebox(0,0){\strut{}\tiny $J$}}%
      \csname LTb\endcsname
      \put(4192,2252){\makebox(0,0)[r]{\strut{}\tiny $n=0.5$}}%
    }%
    \gplbacktext
    \put(0,0){\includegraphics{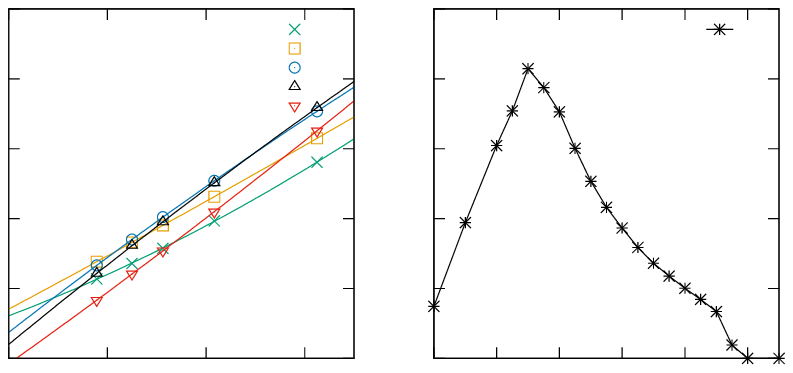}}%
    \gplfronttext
  \end{picture}%
\endgroup

%% file: VEE_Correlations_dimer_spin_peak_L.tex
\begingroup
  \makeatletter
  \providecommand\color[2][]{%
    \GenericError{(gnuplot) \space\space\space\@spaces}{%
      Package color not loaded in conjunction with
      terminal option `colourtext'%
    }{See the gnuplot documentation for explanation.%
    }{Either use 'blacktext' in gnuplot or load the package
      color.sty in LaTeX.}%
    \renewcommand\color[2][]{}%
  }%
  \providecommand\includegraphics[2][]{%
    \GenericError{(gnuplot) \space\space\space\@spaces}{%
      Package graphicx or graphics not loaded%
    }{See the gnuplot documentation for explanation.%
    }{The gnuplot epslatex terminal needs graphicx.sty or graphics.sty.}%
    \renewcommand\includegraphics[2][]{}%
  }%
  \providecommand\rotatebox[2]{#2}%
  \@ifundefined{ifGPcolor}{%
    \newif\ifGPcolor
    \GPcolortrue
  }{}%
  \@ifundefined{ifGPblacktext}{%
    \newif\ifGPblacktext
    \GPblacktexttrue
  }{}%
  \let\gplgaddtomacro\g@addto@macro
  \gdef\gplbacktext{}%
  \gdef\gplfronttext{}%
  \makeatother
  \ifGPblacktext
    \def\colorrgb#1{}%
    \def\colorgray#1{}%
  \else
    \ifGPcolor
      \def\colorrgb#1{\color[rgb]{#1}}%
      \def\colorgray#1{\color[gray]{#1}}%
      \expandafter\def\csname LTw\endcsname{\color{white}}%
      \expandafter\def\csname LTb\endcsname{\color{black}}%
      \expandafter\def\csname LTa\endcsname{\color{black}}%
      \expandafter\def\csname LT0\endcsname{\color[rgb]{1,0,0}}%
      \expandafter\def\csname LT1\endcsname{\color[rgb]{0,1,0}}%
      \expandafter\def\csname LT2\endcsname{\color[rgb]{0,0,1}}%
      \expandafter\def\csname LT3\endcsname{\color[rgb]{1,0,1}}%
      \expandafter\def\csname LT4\endcsname{\color[rgb]{0,1,1}}%
      \expandafter\def\csname LT5\endcsname{\color[rgb]{1,1,0}}%
      \expandafter\def\csname LT6\endcsname{\color[rgb]{0,0,0}}%
      \expandafter\def\csname LT7\endcsname{\color[rgb]{1,0.3,0}}%
      \expandafter\def\csname LT8\endcsname{\color[rgb]{0.5,0.5,0.5}}%
    \else
      \def\colorrgb#1{\color{black}}%
      \def\colorgray#1{\color[gray]{#1}}%
      \expandafter\def\csname LTw\endcsname{\color{white}}%
      \expandafter\def\csname LTb\endcsname{\color{black}}%
      \expandafter\def\csname LTa\endcsname{\color{black}}%
      \expandafter\def\csname LT0\endcsname{\color{black}}%
      \expandafter\def\csname LT1\endcsname{\color{black}}%
      \expandafter\def\csname LT2\endcsname{\color{black}}%
      \expandafter\def\csname LT3\endcsname{\color{black}}%
      \expandafter\def\csname LT4\endcsname{\color{black}}%
      \expandafter\def\csname LT5\endcsname{\color{black}}%
      \expandafter\def\csname LT6\endcsname{\color{black}}%
      \expandafter\def\csname LT7\endcsname{\color{black}}%
      \expandafter\def\csname LT8\endcsname{\color{black}}%
    \fi
  \fi
    \setlength{\unitlength}{0.0500bp}%
    \ifx\gptboxheight\undefined%
      \newlength{\gptboxheight}%
      \newlength{\gptboxwidth}%
      \newsavebox{\gptboxtext}%
    \fi%
    \setlength{\fboxrule}{0.5pt}%
    \setlength{\fboxsep}{1pt}%
\begin{picture}(5102.00,5102.00)%
    \gplgaddtomacro\gplbacktext{%
      \csname LTb\endcsname
      \put(279,2806){\makebox(0,0)[r]{\strut{}\tiny 1.2}}%
      \put(279,3061){\makebox(0,0)[r]{\strut{}\tiny 1.4}}%
      \put(279,3316){\makebox(0,0)[r]{\strut{}\tiny 1.6}}%
      \put(279,3571){\makebox(0,0)[r]{\strut{}\tiny 1.8}}%
      \put(279,3826){\makebox(0,0)[r]{\strut{}\tiny 2}}%
      \put(279,4080){\makebox(0,0)[r]{\strut{}\tiny 2.2}}%
      \put(279,4335){\makebox(0,0)[r]{\strut{}\tiny 2.4}}%
      \put(279,4590){\makebox(0,0)[r]{\strut{}\tiny 2.6}}%
      \put(279,4845){\makebox(0,0)[r]{\strut{}\tiny 2.8}}%
      \put(306,2696){\makebox(0,0){\strut{}\tiny 0}}%
      \put(561,2696){\makebox(0,0){\strut{}\tiny 20}}%
      \put(816,2696){\makebox(0,0){\strut{}\tiny 40}}%
      \put(1071,2696){\makebox(0,0){\strut{}\tiny 60}}%
      \put(1326,2696){\makebox(0,0){\strut{}\tiny 80}}%
      \put(1581,2696){\makebox(0,0){\strut{}\tiny 100}}%
      \put(1836,2696){\makebox(0,0){\strut{}\tiny 120}}%
      \put(2091,2696){\makebox(0,0){\strut{}\tiny 140}}%
      \put(2346,2696){\makebox(0,0){\strut{}\tiny 160}}%
      \put(752,3953){\makebox(0,0)[l]{\strut{}\tiny $c(L=80) = 1.19 \pm 0.09$}}%
      \put(752,3826){\makebox(0,0)[l]{\strut{}\tiny $c(L=112) = 1.07 \pm 0.08$}}%
      \put(752,3698){\makebox(0,0)[l]{\strut{}\tiny $c(L=160) = 0.96 \pm 0.06$}}%
      \put(689,3061){\makebox(0,0)[l]{\strut{}\tiny (a)}}%
    }%
    \gplgaddtomacro\gplfronttext{%
      \csname LTb\endcsname
      \put(-86,3825){\rotatebox{-270}{\makebox(0,0){\strut{}\tiny $S_{EE}(i)$}}}%
      \put(1326,2586){\makebox(0,0){\strut{}\tiny $i$}}%
      \csname LTb\endcsname
      \put(1795,3309){\makebox(0,0)[r]{\strut{}\tiny $L=80$}}%
      \csname LTb\endcsname
      \put(1795,3133){\makebox(0,0)[r]{\strut{}\tiny $L=112$}}%
      \csname LTb\endcsname
      \put(1795,2957){\makebox(0,0)[r]{\strut{}\tiny $L=160$}}%
    }%
    \gplgaddtomacro\gplbacktext{%
      \csname LTb\endcsname
      \put(2830,2806){\makebox(0,0)[r]{\strut{}\tiny -0.4}}%
      \put(2830,3097){\makebox(0,0)[r]{\strut{}\tiny -0.3}}%
      \put(2830,3389){\makebox(0,0)[r]{\strut{}\tiny -0.2}}%
      \put(2830,3680){\makebox(0,0)[r]{\strut{}\tiny -0.1}}%
      \put(2830,3971){\makebox(0,0)[r]{\strut{}\tiny 0}}%
      \put(2830,4262){\makebox(0,0)[r]{\strut{}\tiny 0.1}}%
      \put(2830,4554){\makebox(0,0)[r]{\strut{}\tiny 0.2}}%
      \put(2830,4845){\makebox(0,0)[r]{\strut{}\tiny 0.3}}%
      \put(2857,2696){\makebox(0,0){\strut{}\tiny 0}}%
      \put(3112,2696){\makebox(0,0){\strut{}\tiny 10}}%
      \put(3367,2696){\makebox(0,0){\strut{}\tiny 20}}%
      \put(3622,2696){\makebox(0,0){\strut{}\tiny 30}}%
      \put(3877,2696){\makebox(0,0){\strut{}\tiny 40}}%
      \put(4131,2696){\makebox(0,0){\strut{}\tiny 50}}%
      \put(4386,2696){\makebox(0,0){\strut{}\tiny 60}}%
      \put(4641,2696){\makebox(0,0){\strut{}\tiny 70}}%
      \put(4896,2696){\makebox(0,0){\strut{}\tiny 80}}%
      \put(3214,3097){\makebox(0,0)[l]{\strut{}\tiny (b)}}%
    }%
    \gplgaddtomacro\gplfronttext{%
      \csname LTb\endcsname
      \put(3876,2586){\makebox(0,0){\strut{}\tiny $i$}}%
      \csname LTb\endcsname
      \put(4345,4727){\makebox(0,0)[r]{\strut{}\tiny $< \overrightarrow{S}_{x} \cdot \overrightarrow{S}_{x+i}>$}}%
      \csname LTb\endcsname
      \put(4345,4617){\makebox(0,0)[r]{\strut{}\tiny $<D_{x} D_{x+i}>$}}%
    }%
    \gplgaddtomacro\gplbacktext{%
      \csname LTb\endcsname
      \put(279,408){\makebox(0,0)[r]{\strut{}\tiny $10^{-9}$}}%
      \put(279,635){\makebox(0,0)[r]{\strut{}\tiny $10^{-8}$}}%
      \put(279,861){\makebox(0,0)[r]{\strut{}\tiny $10^{-7}$}}%
      \put(279,1088){\makebox(0,0)[r]{\strut{}\tiny $10^{-6}$}}%
      \put(279,1315){\makebox(0,0)[r]{\strut{}\tiny $10^{-5}$}}%
      \put(279,1541){\makebox(0,0)[r]{\strut{}\tiny $10^{-4}$}}%
      \put(279,1768){\makebox(0,0)[r]{\strut{}\tiny $10^{-3}$}}%
      \put(279,1995){\makebox(0,0)[r]{\strut{}\tiny $10^{-2}$}}%
      \put(279,2221){\makebox(0,0)[r]{\strut{}\tiny $10^{-1}$}}%
      \put(279,2448){\makebox(0,0)[r]{\strut{}\tiny $10^{0}$}}%
      \put(306,298){\makebox(0,0){\strut{}\tiny 0}}%
      \put(561,298){\makebox(0,0){\strut{}\tiny 10}}%
      \put(816,298){\makebox(0,0){\strut{}\tiny 20}}%
      \put(1071,298){\makebox(0,0){\strut{}\tiny 30}}%
      \put(1326,298){\makebox(0,0){\strut{}\tiny 40}}%
      \put(1581,298){\makebox(0,0){\strut{}\tiny 50}}%
      \put(1836,298){\makebox(0,0){\strut{}\tiny 60}}%
      \put(2091,298){\makebox(0,0){\strut{}\tiny 70}}%
      \put(2346,298){\makebox(0,0){\strut{}\tiny 80}}%
      \put(663,635){\makebox(0,0)[l]{\strut{}\tiny (c)}}%
    }%
    \gplgaddtomacro\gplfronttext{%
      \csname LTb\endcsname
      \put(1326,188){\makebox(0,0){\strut{}\tiny $i$}}%
      \csname LTb\endcsname
      \put(1795,1483){\makebox(0,0)[r]{\strut{}\tiny $|<\overrightarrow{S}_{x} \cdot \overrightarrow{S}_{x+i}>|$}}%
      \csname LTb\endcsname
      \put(1795,1373){\makebox(0,0)[r]{\strut{}\tiny $|<D_{x} D_{x+i}>|$}}%
    }%
    \gplgaddtomacro\gplbacktext{%
      \csname LTb\endcsname
      \put(2830,408){\makebox(0,0)[r]{\strut{}\tiny 0}}%
      \put(2830,816){\makebox(0,0)[r]{\strut{}\tiny 0.01}}%
      \put(2830,1224){\makebox(0,0)[r]{\strut{}\tiny 0.02}}%
      \put(2830,1632){\makebox(0,0)[r]{\strut{}\tiny 0.03}}%
      \put(2830,2040){\makebox(0,0)[r]{\strut{}\tiny 0.04}}%
      \put(2830,2448){\makebox(0,0)[r]{\strut{}\tiny 0.05}}%
      \put(2857,298){\makebox(0,0){\strut{}\tiny 0}}%
      \put(3440,298){\makebox(0,0){\strut{}\tiny 0.01}}%
      \put(4022,298){\makebox(0,0){\strut{}\tiny 0.02}}%
      \put(4605,298){\makebox(0,0){\strut{}\tiny 0.03}}%
      \put(3265,816){\makebox(0,0)[l]{\strut{}\tiny (d)}}%
    }%
    \gplgaddtomacro\gplfronttext{%
      \csname LTb\endcsname
      \put(3876,188){\makebox(0,0){\strut{}\tiny $1/L$}}%
      \csname LTb\endcsname
      \put(4345,559){\makebox(0,0)[r]{\strut{}\tiny $S(q=\frac{\pi}{2} ) /L$}}%
    }%
    \gplbacktext
    \put(0,0){\includegraphics{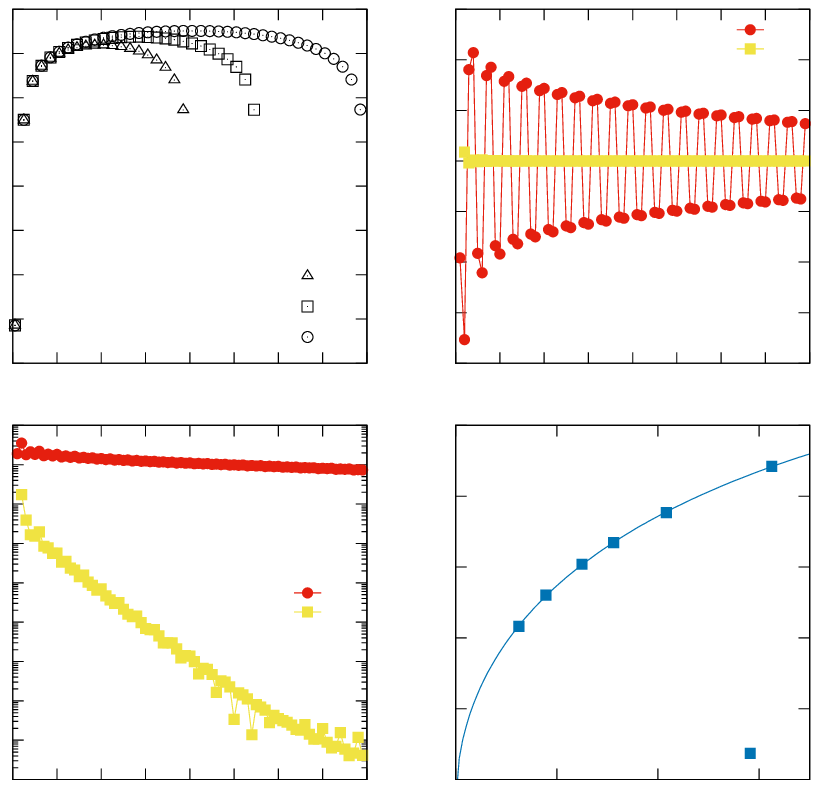}}%
    \gplfronttext
  \end{picture}%
\endgroup

%% file: states_converge.tex
\begingroup
  \makeatletter
  \providecommand\color[2][]{%
    \GenericError{(gnuplot) \space\space\space\@spaces}{%
      Package color not loaded in conjunction with
      terminal option `colourtext'%
    }{See the gnuplot documentation for explanation.%
    }{Either use 'blacktext' in gnuplot or load the package
      color.sty in LaTeX.}%
    \renewcommand\color[2][]{}%
  }%
  \providecommand\includegraphics[2][]{%
    \GenericError{(gnuplot) \space\space\space\@spaces}{%
      Package graphicx or graphics not loaded%
    }{See the gnuplot documentation for explanation.%
    }{The gnuplot epslatex terminal needs graphicx.sty or graphics.sty.}%
    \renewcommand\includegraphics[2][]{}%
  }%
  \providecommand\rotatebox[2]{#2}%
  \@ifundefined{ifGPcolor}{%
    \newif\ifGPcolor
    \GPcolortrue
  }{}%
  \@ifundefined{ifGPblacktext}{%
    \newif\ifGPblacktext
    \GPblacktexttrue
  }{}%
  \let\gplgaddtomacro\g@addto@macro
  \gdef\gplbacktext{}%
  \gdef\gplfronttext{}%
  \makeatother
  \ifGPblacktext
    \def\colorrgb#1{}%
    \def\colorgray#1{}%
  \else
    \ifGPcolor
      \def\colorrgb#1{\color[rgb]{#1}}%
      \def\colorgray#1{\color[gray]{#1}}%
      \expandafter\def\csname LTw\endcsname{\color{white}}%
      \expandafter\def\csname LTb\endcsname{\color{black}}%
      \expandafter\def\csname LTa\endcsname{\color{black}}%
      \expandafter\def\csname LT0\endcsname{\color[rgb]{1,0,0}}%
      \expandafter\def\csname LT1\endcsname{\color[rgb]{0,1,0}}%
      \expandafter\def\csname LT2\endcsname{\color[rgb]{0,0,1}}%
      \expandafter\def\csname LT3\endcsname{\color[rgb]{1,0,1}}%
      \expandafter\def\csname LT4\endcsname{\color[rgb]{0,1,1}}%
      \expandafter\def\csname LT5\endcsname{\color[rgb]{1,1,0}}%
      \expandafter\def\csname LT6\endcsname{\color[rgb]{0,0,0}}%
      \expandafter\def\csname LT7\endcsname{\color[rgb]{1,0.3,0}}%
      \expandafter\def\csname LT8\endcsname{\color[rgb]{0.5,0.5,0.5}}%
    \else
      \def\colorrgb#1{\color{black}}%
      \def\colorgray#1{\color[gray]{#1}}%
      \expandafter\def\csname LTw\endcsname{\color{white}}%
      \expandafter\def\csname LTb\endcsname{\color{black}}%
      \expandafter\def\csname LTa\endcsname{\color{black}}%
      \expandafter\def\csname LT0\endcsname{\color{black}}%
      \expandafter\def\csname LT1\endcsname{\color{black}}%
      \expandafter\def\csname LT2\endcsname{\color{black}}%
      \expandafter\def\csname LT3\endcsname{\color{black}}%
      \expandafter\def\csname LT4\endcsname{\color{black}}%
      \expandafter\def\csname LT5\endcsname{\color{black}}%
      \expandafter\def\csname LT6\endcsname{\color{black}}%
      \expandafter\def\csname LT7\endcsname{\color{black}}%
      \expandafter\def\csname LT8\endcsname{\color{black}}%
    \fi
  \fi
    \setlength{\unitlength}{0.0500bp}%
    \ifx\gptboxheight\undefined%
      \newlength{\gptboxheight}%
      \newlength{\gptboxwidth}%
      \newsavebox{\gptboxtext}%
    \fi%
    \setlength{\fboxrule}{0.5pt}%
    \setlength{\fboxsep}{1pt}%
\begin{picture}(5102.00,5102.00)%
    \gplgaddtomacro\gplbacktext{%
      \csname LTb\endcsname
      \put(585,3061){\makebox(0,0)[r]{\strut{}\tiny -104.13}}%
      \put(585,3507){\makebox(0,0)[r]{\strut{}\tiny -104.1295}}%
      \put(585,3953){\makebox(0,0)[r]{\strut{}\tiny -104.129}}%
      \put(585,4399){\makebox(0,0)[r]{\strut{}\tiny -104.1285}}%
      \put(585,4845){\makebox(0,0)[r]{\strut{}\tiny -104.128}}%
      \put(835,2951){\makebox(0,0){\strut{}\tiny 5000}}%
      \put(1281,2951){\makebox(0,0){\strut{}\tiny 6000}}%
      \put(1728,2951){\makebox(0,0){\strut{}\tiny 7000}}%
      \put(2174,2951){\makebox(0,0){\strut{}\tiny 8000}}%
      \put(835,3373){\makebox(0,0)[l]{\strut{}\tiny (a)}}%
    }%
    \gplgaddtomacro\gplfronttext{%
      \csname LTb\endcsname
      \put(-110,3953){\rotatebox{-270}{\makebox(0,0){\strut{}\tiny $ E_{0} $}}}%
      \put(1504,2678){\makebox(0,0){\strut{}\tiny $ states $}}%
      \csname LTb\endcsname
      \put(1846,4694){\makebox(0,0)[r]{\strut{}\tiny $ L=112, J=0.6 $}}%
    }%
    \gplgaddtomacro\gplbacktext{%
      \csname LTb\endcsname
      \put(3034,3061){\makebox(0,0)[r]{\strut{}\tiny 0.19}}%
      \put(3034,3418){\makebox(0,0)[r]{\strut{}\tiny 0.191}}%
      \put(3034,3775){\makebox(0,0)[r]{\strut{}\tiny 0.192}}%
      \put(3034,4131){\makebox(0,0)[r]{\strut{}\tiny 0.193}}%
      \put(3034,4488){\makebox(0,0)[r]{\strut{}\tiny 0.194}}%
      \put(3034,4845){\makebox(0,0)[r]{\strut{}\tiny 0.195}}%
      \put(3284,2951){\makebox(0,0){\strut{}\tiny 5000}}%
      \put(3730,2951){\makebox(0,0){\strut{}\tiny 6000}}%
      \put(4176,2951){\makebox(0,0){\strut{}\tiny 7000}}%
      \put(4622,2951){\makebox(0,0){\strut{}\tiny 8000}}%
      \put(3284,3418){\makebox(0,0)[l]{\strut{}\tiny (b)}}%
    }%
    \gplgaddtomacro\gplfronttext{%
      \csname LTb\endcsname
      \put(2537,3953){\rotatebox{-270}{\makebox(0,0){\strut{}\tiny $ \left \langle D(L) \right \rangle$}}}%
      \put(3953,2678){\makebox(0,0){\strut{}\tiny $ states $}}%
      \csname LTb\endcsname
      \put(4294,4694){\makebox(0,0)[r]{\strut{}\tiny $ L=112, J=0.6 $}}%
    }%
    \gplgaddtomacro\gplbacktext{%
      \csname LTb\endcsname
      \put(483,510){\makebox(0,0)[r]{\strut{}\tiny 0.0098}}%
      \put(483,867){\makebox(0,0)[r]{\strut{}\tiny 0.0099}}%
      \put(483,1224){\makebox(0,0)[r]{\strut{}\tiny 0.01}}%
      \put(483,1581){\makebox(0,0)[r]{\strut{}\tiny 0.0101}}%
      \put(483,1938){\makebox(0,0)[r]{\strut{}\tiny 0.0102}}%
      \put(483,2295){\makebox(0,0)[r]{\strut{}\tiny 0.0103}}%
      \put(733,400){\makebox(0,0){\strut{}\tiny 5000}}%
      \put(1179,400){\makebox(0,0){\strut{}\tiny 6000}}%
      \put(1626,400){\makebox(0,0){\strut{}\tiny 7000}}%
      \put(2072,400){\makebox(0,0){\strut{}\tiny 8000}}%
      \put(733,867){\makebox(0,0)[l]{\strut{}\tiny (c)}}%
    }%
    \gplgaddtomacro\gplfronttext{%
      \csname LTb\endcsname
      \put(-106,1402){\rotatebox{-270}{\makebox(0,0){\strut{}\tiny $ \left \langle B(L) \right \rangle $}}}%
      \put(1402,128){\makebox(0,0){\strut{}\tiny $ states $}}%
      \csname LTb\endcsname
      \put(1744,2144){\makebox(0,0)[r]{\strut{}\tiny $ L=112, J=0.6 $}}%
    }%
    \gplgaddtomacro\gplbacktext{%
      \csname LTb\endcsname
      \put(3034,510){\makebox(0,0)[r]{\strut{}\tiny $10^{-4}$}}%
      \put(3034,1403){\makebox(0,0)[r]{\strut{}\tiny $10^{-3}$}}%
      \put(3034,2295){\makebox(0,0)[r]{\strut{}\tiny $10^{-2}$}}%
      \put(3284,400){\makebox(0,0){\strut{}\tiny 5000}}%
      \put(3730,400){\makebox(0,0){\strut{}\tiny 6000}}%
      \put(4176,400){\makebox(0,0){\strut{}\tiny 7000}}%
      \put(4622,400){\makebox(0,0){\strut{}\tiny 8000}}%
      \put(3284,779){\makebox(0,0)[l]{\strut{}\tiny (d)}}%
    }%
    \gplgaddtomacro\gplfronttext{%
      \csname LTb\endcsname
      \put(2577,1402){\rotatebox{-270}{\makebox(0,0){\strut{}\tiny $ \left \langle S^{z}_{L/2} \right \rangle $}}}%
      \put(3953,128){\makebox(0,0){\strut{}\tiny $ states $}}%
      \csname LTb\endcsname
      \put(4294,2144){\makebox(0,0)[r]{\strut{}\tiny $ L=112, J=0.6 $}}%
    }%
    \gplbacktext
    \put(0,0){\includegraphics{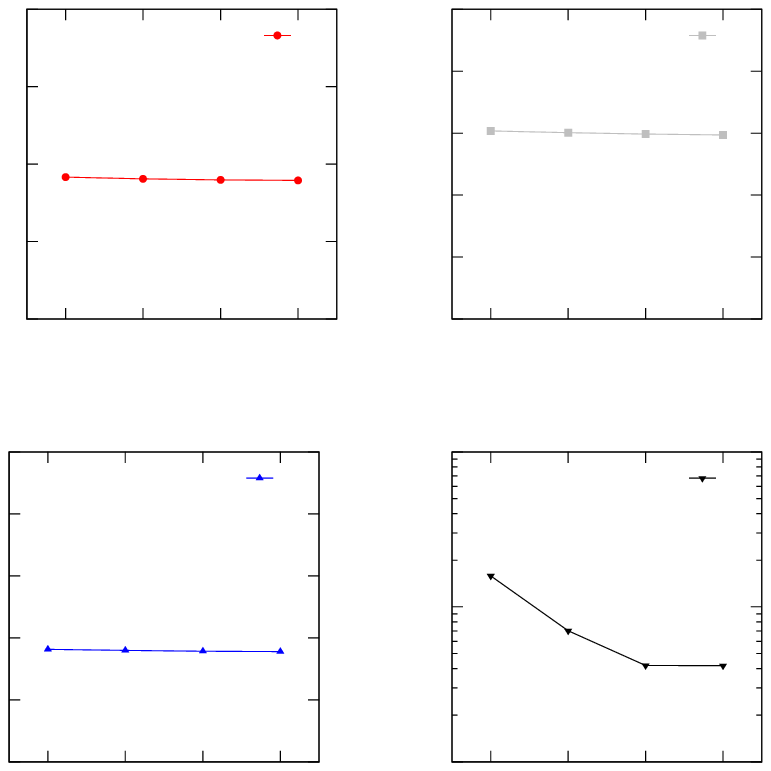}}%
    \gplfronttext
  \end{picture}%
\endgroup

%% file: iDMRG_energy_j08.tex
\begingroup
  \makeatletter
  \providecommand\color[2][]{%
    \GenericError{(gnuplot) \space\space\space\@spaces}{%
      Package color not loaded in conjunction with
      terminal option `colourtext'%
    }{See the gnuplot documentation for explanation.%
    }{Either use 'blacktext' in gnuplot or load the package
      color.sty in LaTeX.}%
    \renewcommand\color[2][]{}%
  }%
  \providecommand\includegraphics[2][]{%
    \GenericError{(gnuplot) \space\space\space\@spaces}{%
      Package graphicx or graphics not loaded%
    }{See the gnuplot documentation for explanation.%
    }{The gnuplot epslatex terminal needs graphicx.sty or graphics.sty.}%
    \renewcommand\includegraphics[2][]{}%
  }%
  \providecommand\rotatebox[2]{#2}%
  \@ifundefined{ifGPcolor}{%
    \newif\ifGPcolor
    \GPcolortrue
  }{}%
  \@ifundefined{ifGPblacktext}{%
    \newif\ifGPblacktext
    \GPblacktexttrue
  }{}%
  \let\gplgaddtomacro\g@addto@macro
  \gdef\gplbacktext{}%
  \gdef\gplfronttext{}%
  \makeatother
  \ifGPblacktext
    \def\colorrgb#1{}%
    \def\colorgray#1{}%
  \else
    \ifGPcolor
      \def\colorrgb#1{\color[rgb]{#1}}%
      \def\colorgray#1{\color[gray]{#1}}%
      \expandafter\def\csname LTw\endcsname{\color{white}}%
      \expandafter\def\csname LTb\endcsname{\color{black}}%
      \expandafter\def\csname LTa\endcsname{\color{black}}%
      \expandafter\def\csname LT0\endcsname{\color[rgb]{1,0,0}}%
      \expandafter\def\csname LT1\endcsname{\color[rgb]{0,1,0}}%
      \expandafter\def\csname LT2\endcsname{\color[rgb]{0,0,1}}%
      \expandafter\def\csname LT3\endcsname{\color[rgb]{1,0,1}}%
      \expandafter\def\csname LT4\endcsname{\color[rgb]{0,1,1}}%
      \expandafter\def\csname LT5\endcsname{\color[rgb]{1,1,0}}%
      \expandafter\def\csname LT6\endcsname{\color[rgb]{0,0,0}}%
      \expandafter\def\csname LT7\endcsname{\color[rgb]{1,0.3,0}}%
      \expandafter\def\csname LT8\endcsname{\color[rgb]{0.5,0.5,0.5}}%
    \else
      \def\colorrgb#1{\color{black}}%
      \def\colorgray#1{\color[gray]{#1}}%
      \expandafter\def\csname LTw\endcsname{\color{white}}%
      \expandafter\def\csname LTb\endcsname{\color{black}}%
      \expandafter\def\csname LTa\endcsname{\color{black}}%
      \expandafter\def\csname LT0\endcsname{\color{black}}%
      \expandafter\def\csname LT1\endcsname{\color{black}}%
      \expandafter\def\csname LT2\endcsname{\color{black}}%
      \expandafter\def\csname LT3\endcsname{\color{black}}%
      \expandafter\def\csname LT4\endcsname{\color{black}}%
      \expandafter\def\csname LT5\endcsname{\color{black}}%
      \expandafter\def\csname LT6\endcsname{\color{black}}%
      \expandafter\def\csname LT7\endcsname{\color{black}}%
      \expandafter\def\csname LT8\endcsname{\color{black}}%
    \fi
  \fi
    \setlength{\unitlength}{0.0500bp}%
    \ifx\gptboxheight\undefined%
      \newlength{\gptboxheight}%
      \newlength{\gptboxwidth}%
      \newsavebox{\gptboxtext}%
    \fi%
    \setlength{\fboxrule}{0.5pt}%
    \setlength{\fboxsep}{1pt}%
\begin{picture}(3968.00,3968.00)%
    \gplgaddtomacro\gplbacktext{%
      \csname LTb\endcsname
      \put(264,396){\makebox(0,0)[r]{\strut{}\tiny -0.97}}%
      \put(264,1190){\makebox(0,0)[r]{\strut{}\tiny -0.965}}%
      \put(264,1983){\makebox(0,0)[r]{\strut{}\tiny -0.96}}%
      \put(264,2777){\makebox(0,0)[r]{\strut{}\tiny -0.955}}%
      \put(264,3570){\makebox(0,0)[r]{\strut{}\tiny -0.95}}%
      \put(396,176){\makebox(0,0){\strut{}\tiny 4000}}%
      \put(973,176){\makebox(0,0){\strut{}\tiny 6000}}%
      \put(1550,176){\makebox(0,0){\strut{}\tiny 8000}}%
      \put(2127,176){\makebox(0,0){\strut{}\tiny 10000}}%
      \put(2704,176){\makebox(0,0){\strut{}\tiny 12000}}%
      \put(3281,176){\makebox(0,0){\strut{}\tiny 14000}}%
    }%
    \gplgaddtomacro\gplfronttext{%
      \csname LTb\endcsname
      \put(-392,1983){\rotatebox{-270}{\makebox(0,0){\strut{}\tiny energy per site}}}%
      \put(1983,0){\makebox(0,0){\strut{}\tiny states}}%
      \csname LTb\endcsname
      \put(2583,3397){\makebox(0,0)[r]{\strut{}\tiny iDMRG}}%
      \csname LTb\endcsname
      \put(2583,3177){\makebox(0,0)[r]{\strut{}\tiny finite size DMRG with L=112}}%
    }%
    \gplbacktext
    \put(0,0){\includegraphics{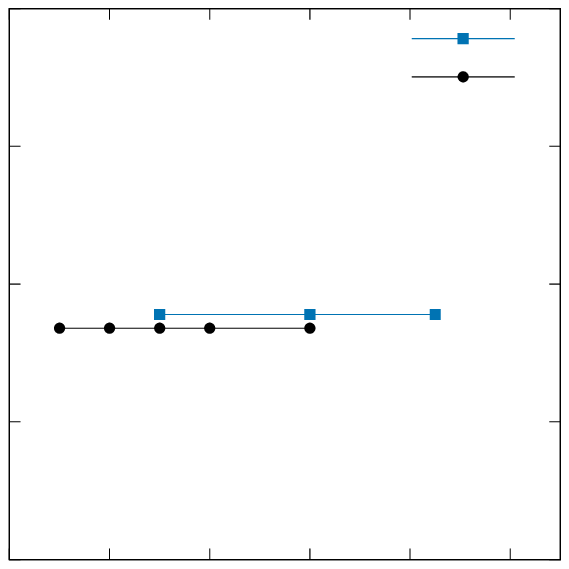}}%
    \gplfronttext
  \end{picture}%
\endgroup

%% file: KLM_BOW.bbl
\begin{thebibliography}{48}%
\makeatletter
\providecommand \@ifxundefined [1]{%
 \@ifx{#1\undefined}
}%
\providecommand \@ifnum [1]{%
 \ifnum #1\expandafter \@firstoftwo
 \else \expandafter \@secondoftwo
 \fi
}%
\providecommand \@ifx [1]{%
 \ifx #1\expandafter \@firstoftwo
 \else \expandafter \@secondoftwo
 \fi
}%
\providecommand \natexlab [1]{#1}%
\providecommand \enquote  [1]{``#1''}%
\providecommand \bibnamefont  [1]{#1}%
\providecommand \bibfnamefont [1]{#1}%
\providecommand \citenamefont [1]{#1}%
\providecommand \href@noop [0]{\@secondoftwo}%
\providecommand \href [0]{\begingroup \@sanitize@url \@href}%
\providecommand \@href[1]{\@@startlink{#1}\@@href}%
\providecommand \@@href[1]{\endgroup#1\@@endlink}%
\providecommand \@sanitize@url [0]{\catcode `\\12\catcode `\$12\catcode
  `\&12\catcode `\#12\catcode `\^12\catcode `\_12\catcode `\%12\relax}%
\providecommand \@@startlink[1]{}%
\providecommand \@@endlink[0]{}%
\providecommand \url  [0]{\begingroup\@sanitize@url \@url }%
\providecommand \@url [1]{\endgroup\@href {#1}{\urlprefix }}%
\providecommand \urlprefix  [0]{URL }%
\providecommand \Eprint [0]{\href }%
\providecommand \doibase [0]{http://dx.doi.org/}%
\providecommand \selectlanguage [0]{\@gobble}%
\providecommand \bibinfo  [0]{\@secondoftwo}%
\providecommand \bibfield  [0]{\@secondoftwo}%
\providecommand \translation [1]{[#1]}%
\providecommand \BibitemOpen [0]{}%
\providecommand \bibitemStop [0]{}%
\providecommand \bibitemNoStop [0]{.\EOS\space}%
\providecommand \EOS [0]{\spacefactor3000\relax}%
\providecommand \BibitemShut  [1]{\csname bibitem#1\endcsname}%
\let\auto@bib@innerbib\@empty
\bibitem [{\citenamefont {Comes}\ \emph {et~al.}(1973)\citenamefont {Comes},
  \citenamefont {Lambert}, \citenamefont {Launois},\ and\ \citenamefont
  {Zeller}}]{comes1973evidence}%
  \BibitemOpen
  \bibfield  {author} {\bibinfo {author} {\bibfnamefont {R.}~\bibnamefont
  {Comes}}, \bibinfo {author} {\bibfnamefont {M.}~\bibnamefont {Lambert}},
  \bibinfo {author} {\bibfnamefont {H.}~\bibnamefont {Launois}}, \ and\
  \bibinfo {author} {\bibfnamefont {H.}~\bibnamefont {Zeller}},\ }\href@noop {}
  {\bibfield  {journal} {\bibinfo  {journal} {Physical Review B}\ }\textbf
  {\bibinfo {volume} {8}},\ \bibinfo {pages} {571} (\bibinfo {year}
  {1973})}\BibitemShut {NoStop}%
\bibitem [{\citenamefont {Pouget}(2012)}]{pouget2012bond}%
  \BibitemOpen
  \bibfield  {author} {\bibinfo {author} {\bibfnamefont {J.-P.}\ \bibnamefont
  {Pouget}},\ }\href@noop {} {\bibfield  {journal} {\bibinfo  {journal}
  {Physica B: Condensed Matter}\ }\textbf {\bibinfo {volume} {407}},\ \bibinfo
  {pages} {1762} (\bibinfo {year} {2012})}\BibitemShut {NoStop}%
\bibitem [{\citenamefont {Travaglini}\ and\ \citenamefont
  {Wachter}(1984)}]{travaglini1984charge}%
  \BibitemOpen
  \bibfield  {author} {\bibinfo {author} {\bibfnamefont {G.}~\bibnamefont
  {Travaglini}}\ and\ \bibinfo {author} {\bibfnamefont {P.}~\bibnamefont
  {Wachter}},\ }\href@noop {} {\bibfield  {journal} {\bibinfo  {journal}
  {Physical Review B}\ }\textbf {\bibinfo {volume} {30}},\ \bibinfo {pages}
  {1971} (\bibinfo {year} {1984})}\BibitemShut {NoStop}%
\bibitem [{\citenamefont {Pouget}\ \emph {et~al.}(1991)\citenamefont {Pouget},
  \citenamefont {Hennion}, \citenamefont {Escribe-Filippini},\ and\
  \citenamefont {Sato}}]{pouget1991neutron}%
  \BibitemOpen
  \bibfield  {author} {\bibinfo {author} {\bibfnamefont {J.}~\bibnamefont
  {Pouget}}, \bibinfo {author} {\bibfnamefont {B.}~\bibnamefont {Hennion}},
  \bibinfo {author} {\bibfnamefont {C.}~\bibnamefont {Escribe-Filippini}}, \
  and\ \bibinfo {author} {\bibfnamefont {M.}~\bibnamefont {Sato}},\ }\href@noop
  {} {\bibfield  {journal} {\bibinfo  {journal} {Physical Review B}\ }\textbf
  {\bibinfo {volume} {43}},\ \bibinfo {pages} {8421} (\bibinfo {year}
  {1991})}\BibitemShut {NoStop}%
\bibitem [{\citenamefont {Schlenker}\ \emph {et~al.}(1996)\citenamefont
  {Schlenker}, \citenamefont {Dumas}, \citenamefont {Greenblatt},\ and\
  \citenamefont {van Smaalen}}]{schlenker1996physics}%
  \BibitemOpen
  \bibfield  {author} {\bibinfo {author} {\bibfnamefont {C.}~\bibnamefont
  {Schlenker}}, \bibinfo {author} {\bibfnamefont {J.}~\bibnamefont {Dumas}},
  \bibinfo {author} {\bibfnamefont {M.}~\bibnamefont {Greenblatt}}, \ and\
  \bibinfo {author} {\bibfnamefont {S.}~\bibnamefont {van Smaalen}},\
  }\href@noop {} {\emph {\bibinfo {title} {Physics and Chemistry of
  Low-Dimensional Inorganic Conductors}}},\ Vol.\ \bibinfo {volume} {354}\
  (\bibinfo  {publisher} {Springer Science \& Business Media},\ \bibinfo {year}
  {1996})\BibitemShut {NoStop}%
\bibitem [{\citenamefont {Hase}\ \emph
  {et~al.}(1993{\natexlab{a}})\citenamefont {Hase}, \citenamefont {Terasaki},\
  and\ \citenamefont {Uchinokura}}]{hase1993observation}%
  \BibitemOpen
  \bibfield  {author} {\bibinfo {author} {\bibfnamefont {M.}~\bibnamefont
  {Hase}}, \bibinfo {author} {\bibfnamefont {I.}~\bibnamefont {Terasaki}}, \
  and\ \bibinfo {author} {\bibfnamefont {K.}~\bibnamefont {Uchinokura}},\
  }\href@noop {} {\bibfield  {journal} {\bibinfo  {journal} {Physical Review
  Letters}\ }\textbf {\bibinfo {volume} {70}},\ \bibinfo {pages} {3651}
  (\bibinfo {year} {1993}{\natexlab{a}})}\BibitemShut {NoStop}%
\bibitem [{\citenamefont {Hase}\ \emph
  {et~al.}(1993{\natexlab{b}})\citenamefont {Hase}, \citenamefont {Terasaki},
  \citenamefont {Uchinokura}, \citenamefont {Tokunaga}, \citenamefont {Miura},\
  and\ \citenamefont {Obara}}]{hase1993magnetic}%
  \BibitemOpen
  \bibfield  {author} {\bibinfo {author} {\bibfnamefont {M.}~\bibnamefont
  {Hase}}, \bibinfo {author} {\bibfnamefont {I.}~\bibnamefont {Terasaki}},
  \bibinfo {author} {\bibfnamefont {K.}~\bibnamefont {Uchinokura}}, \bibinfo
  {author} {\bibfnamefont {M.}~\bibnamefont {Tokunaga}}, \bibinfo {author}
  {\bibfnamefont {N.}~\bibnamefont {Miura}}, \ and\ \bibinfo {author}
  {\bibfnamefont {H.}~\bibnamefont {Obara}},\ }\href@noop {} {\bibfield
  {journal} {\bibinfo  {journal} {Physical Review B}\ }\textbf {\bibinfo
  {volume} {48}},\ \bibinfo {pages} {9616} (\bibinfo {year}
  {1993}{\natexlab{b}})}\BibitemShut {NoStop}%
\bibitem [{\citenamefont {Pouget}\ \emph {et~al.}(2017)\citenamefont {Pouget},
  \citenamefont {Foury-Leylekian},\ and\ \citenamefont
  {Almeida}}]{pouget2017peierls}%
  \BibitemOpen
  \bibfield  {author} {\bibinfo {author} {\bibfnamefont {J.-P.}\ \bibnamefont
  {Pouget}}, \bibinfo {author} {\bibfnamefont {P.}~\bibnamefont
  {Foury-Leylekian}}, \ and\ \bibinfo {author} {\bibfnamefont {M.}~\bibnamefont
  {Almeida}},\ }\href@noop {} {\bibfield  {journal} {\bibinfo  {journal}
  {Magnetochemistry}\ }\textbf {\bibinfo {volume} {3}},\ \bibinfo {pages} {13}
  (\bibinfo {year} {2017})}\BibitemShut {NoStop}%
\bibitem [{\citenamefont {Henriques}\ \emph {et~al.}(1984)\citenamefont
  {Henriques}, \citenamefont {Alcacer}, \citenamefont {Pouget},\ and\
  \citenamefont {J{\'e}rome}}]{henriques1984electrical}%
  \BibitemOpen
  \bibfield  {author} {\bibinfo {author} {\bibfnamefont {R.}~\bibnamefont
  {Henriques}}, \bibinfo {author} {\bibfnamefont {L.}~\bibnamefont {Alcacer}},
  \bibinfo {author} {\bibfnamefont {J.}~\bibnamefont {Pouget}}, \ and\ \bibinfo
  {author} {\bibfnamefont {D.}~\bibnamefont {J{\'e}rome}},\ }\href@noop {}
  {\bibfield  {journal} {\bibinfo  {journal} {Journal of Physics C: Solid State
  Physics}\ }\textbf {\bibinfo {volume} {17}},\ \bibinfo {pages} {5197}
  (\bibinfo {year} {1984})}\BibitemShut {NoStop}%
\bibitem [{\citenamefont {Henriques}\ \emph {et~al.}(1986)\citenamefont
  {Henriques}, \citenamefont {Alc{\'a}cer}, \citenamefont {Jerome},
  \citenamefont {Bourbonnais},\ and\ \citenamefont
  {Weyl}}]{henriques1986electrical}%
  \BibitemOpen
  \bibfield  {author} {\bibinfo {author} {\bibfnamefont {R.}~\bibnamefont
  {Henriques}}, \bibinfo {author} {\bibfnamefont {L.}~\bibnamefont
  {Alc{\'a}cer}}, \bibinfo {author} {\bibfnamefont {D.}~\bibnamefont {Jerome}},
  \bibinfo {author} {\bibfnamefont {C.}~\bibnamefont {Bourbonnais}}, \ and\
  \bibinfo {author} {\bibfnamefont {C.}~\bibnamefont {Weyl}},\ }\href@noop {}
  {\bibfield  {journal} {\bibinfo  {journal} {Journal of Physics C: Solid State
  Physics}\ }\textbf {\bibinfo {volume} {19}},\ \bibinfo {pages} {4663}
  (\bibinfo {year} {1986})}\BibitemShut {NoStop}%
\bibitem [{\citenamefont {Bourbonnais}\ \emph {et~al.}(1991)\citenamefont
  {Bourbonnais}, \citenamefont {Henriques}, \citenamefont {Wzietek},
  \citenamefont {K{\"o}ngeter}, \citenamefont {Voiron},\ and\ \citenamefont
  {J{\'e}rme}}]{bourbonnais1991nuclear}%
  \BibitemOpen
  \bibfield  {author} {\bibinfo {author} {\bibfnamefont {C.}~\bibnamefont
  {Bourbonnais}}, \bibinfo {author} {\bibfnamefont {R.}~\bibnamefont
  {Henriques}}, \bibinfo {author} {\bibfnamefont {P.}~\bibnamefont {Wzietek}},
  \bibinfo {author} {\bibfnamefont {D.}~\bibnamefont {K{\"o}ngeter}}, \bibinfo
  {author} {\bibfnamefont {J.}~\bibnamefont {Voiron}}, \ and\ \bibinfo {author}
  {\bibfnamefont {D.}~\bibnamefont {J{\'e}rme}},\ }\href@noop {} {\bibfield
  {journal} {\bibinfo  {journal} {Physical Review B}\ }\textbf {\bibinfo
  {volume} {44}},\ \bibinfo {pages} {641} (\bibinfo {year} {1991})}\BibitemShut
  {NoStop}%
\bibitem [{\citenamefont {Matos}\ \emph {et~al.}(1996)\citenamefont {Matos},
  \citenamefont {Bonfait}, \citenamefont {Henriques},\ and\ \citenamefont
  {Almeida}}]{matos1996modification}%
  \BibitemOpen
  \bibfield  {author} {\bibinfo {author} {\bibfnamefont {M.}~\bibnamefont
  {Matos}}, \bibinfo {author} {\bibfnamefont {G.}~\bibnamefont {Bonfait}},
  \bibinfo {author} {\bibfnamefont {R.~T.}\ \bibnamefont {Henriques}}, \ and\
  \bibinfo {author} {\bibfnamefont {M.}~\bibnamefont {Almeida}},\ }\href@noop
  {} {\bibfield  {journal} {\bibinfo  {journal} {Physical Review B}\ }\textbf
  {\bibinfo {volume} {54}},\ \bibinfo {pages} {15307} (\bibinfo {year}
  {1996})}\BibitemShut {NoStop}%
\bibitem [{\citenamefont {Green}\ \emph {et~al.}(2011)\citenamefont {Green},
  \citenamefont {Brooks}, \citenamefont {Kuhns}, \citenamefont {Reyes},
  \citenamefont {Lumata}, \citenamefont {Almeida}, \citenamefont {Matos},
  \citenamefont {Henriques}, \citenamefont {Wright},\ and\ \citenamefont
  {Brown}}]{green2011interaction}%
  \BibitemOpen
  \bibfield  {author} {\bibinfo {author} {\bibfnamefont {E.}~\bibnamefont
  {Green}}, \bibinfo {author} {\bibfnamefont {J.}~\bibnamefont {Brooks}},
  \bibinfo {author} {\bibfnamefont {P.}~\bibnamefont {Kuhns}}, \bibinfo
  {author} {\bibfnamefont {A.}~\bibnamefont {Reyes}}, \bibinfo {author}
  {\bibfnamefont {L.}~\bibnamefont {Lumata}}, \bibinfo {author} {\bibfnamefont
  {M.}~\bibnamefont {Almeida}}, \bibinfo {author} {\bibfnamefont
  {M.}~\bibnamefont {Matos}}, \bibinfo {author} {\bibfnamefont
  {R.}~\bibnamefont {Henriques}}, \bibinfo {author} {\bibfnamefont
  {J.}~\bibnamefont {Wright}}, \ and\ \bibinfo {author} {\bibfnamefont
  {S.}~\bibnamefont {Brown}},\ }\href@noop {} {\bibfield  {journal} {\bibinfo
  {journal} {Physical Review B}\ }\textbf {\bibinfo {volume} {84}},\ \bibinfo
  {pages} {121101} (\bibinfo {year} {2011})}\BibitemShut {NoStop}%
\bibitem [{\citenamefont {Doniach}(1977)}]{doniach1977kondo}%
  \BibitemOpen
  \bibfield  {author} {\bibinfo {author} {\bibfnamefont {S.}~\bibnamefont
  {Doniach}},\ }\href@noop {} {\bibfield  {journal} {\bibinfo  {journal}
  {physica B+ C}\ }\textbf {\bibinfo {volume} {91}},\ \bibinfo {pages} {231}
  (\bibinfo {year} {1977})}\BibitemShut {NoStop}%
\bibitem [{\citenamefont {Lacroix}\ and\ \citenamefont
  {Cyrot}(1979)}]{lacroix1979phase}%
  \BibitemOpen
  \bibfield  {author} {\bibinfo {author} {\bibfnamefont {C.}~\bibnamefont
  {Lacroix}}\ and\ \bibinfo {author} {\bibfnamefont {M.}~\bibnamefont
  {Cyrot}},\ }\href@noop {} {\bibfield  {journal} {\bibinfo  {journal}
  {Physical Review B}\ }\textbf {\bibinfo {volume} {20}},\ \bibinfo {pages}
  {1969} (\bibinfo {year} {1979})}\BibitemShut {NoStop}%
\bibitem [{\citenamefont {Fazekas}\ and\ \citenamefont
  {M{\"u}ller-Hartmann}(1991)}]{fazekas1991magnetic}%
  \BibitemOpen
  \bibfield  {author} {\bibinfo {author} {\bibfnamefont {P.}~\bibnamefont
  {Fazekas}}\ and\ \bibinfo {author} {\bibfnamefont {E.}~\bibnamefont
  {M{\"u}ller-Hartmann}},\ }\href@noop {} {\bibfield  {journal} {\bibinfo
  {journal} {Zeitschrift f{\"u}r Physik B Condensed Matter}\ }\textbf {\bibinfo
  {volume} {85}},\ \bibinfo {pages} {285} (\bibinfo {year} {1991})}\BibitemShut
  {NoStop}%
\bibitem [{\citenamefont {Tsunetsugu}\ \emph {et~al.}(1993)\citenamefont
  {Tsunetsugu}, \citenamefont {Sigrist},\ and\ \citenamefont
  {Ueda}}]{tsunetsugu1993phase}%
  \BibitemOpen
  \bibfield  {author} {\bibinfo {author} {\bibfnamefont {H.}~\bibnamefont
  {Tsunetsugu}}, \bibinfo {author} {\bibfnamefont {M.}~\bibnamefont {Sigrist}},
  \ and\ \bibinfo {author} {\bibfnamefont {K.}~\bibnamefont {Ueda}},\
  }\href@noop {} {\bibfield  {journal} {\bibinfo  {journal} {Physical Review
  B}\ }\textbf {\bibinfo {volume} {47}},\ \bibinfo {pages} {8345} (\bibinfo
  {year} {1993})}\BibitemShut {NoStop}%
\bibitem [{\citenamefont {Tsunetsugu}\ \emph {et~al.}(1997)\citenamefont
  {Tsunetsugu}, \citenamefont {Sigrist},\ and\ \citenamefont
  {Ueda}}]{tsunetsugu1997ground}%
  \BibitemOpen
  \bibfield  {author} {\bibinfo {author} {\bibfnamefont {H.}~\bibnamefont
  {Tsunetsugu}}, \bibinfo {author} {\bibfnamefont {M.}~\bibnamefont {Sigrist}},
  \ and\ \bibinfo {author} {\bibfnamefont {K.}~\bibnamefont {Ueda}},\
  }\href@noop {} {\bibfield  {journal} {\bibinfo  {journal} {Reviews of Modern
  Physics}\ }\textbf {\bibinfo {volume} {69}},\ \bibinfo {pages} {809}
  (\bibinfo {year} {1997})}\BibitemShut {NoStop}%
\bibitem [{\citenamefont {McCulloch}\ \emph {et~al.}(2002)\citenamefont
  {McCulloch}, \citenamefont {Juozapavicius}, \citenamefont {Rosengren},\ and\
  \citenamefont {Gulacsi}}]{mcculloch2002localized}%
  \BibitemOpen
  \bibfield  {author} {\bibinfo {author} {\bibfnamefont {I.~P.}\ \bibnamefont
  {McCulloch}}, \bibinfo {author} {\bibfnamefont {A.}~\bibnamefont
  {Juozapavicius}}, \bibinfo {author} {\bibfnamefont {A.}~\bibnamefont
  {Rosengren}}, \ and\ \bibinfo {author} {\bibfnamefont {M.}~\bibnamefont
  {Gulacsi}},\ }\href@noop {} {\bibfield  {journal} {\bibinfo  {journal}
  {Physical Review B}\ }\textbf {\bibinfo {volume} {65}},\ \bibinfo {pages}
  {052410} (\bibinfo {year} {2002})}\BibitemShut {NoStop}%
\bibitem [{\citenamefont {Peters}\ and\ \citenamefont
  {Kawakami}(2012)}]{peters2012ferromagnetic}%
  \BibitemOpen
  \bibfield  {author} {\bibinfo {author} {\bibfnamefont {R.}~\bibnamefont
  {Peters}}\ and\ \bibinfo {author} {\bibfnamefont {N.}~\bibnamefont
  {Kawakami}},\ }\href@noop {} {\bibfield  {journal} {\bibinfo  {journal}
  {Physical Review B}\ }\textbf {\bibinfo {volume} {86}},\ \bibinfo {pages}
  {165107} (\bibinfo {year} {2012})}\BibitemShut {NoStop}%
\bibitem [{\citenamefont {Shibata}\ \emph {et~al.}(1996)\citenamefont
  {Shibata}, \citenamefont {Ueda}, \citenamefont {Nishino},\ and\ \citenamefont
  {Ishii}}]{shibata1996friedel}%
  \BibitemOpen
  \bibfield  {author} {\bibinfo {author} {\bibfnamefont {N.}~\bibnamefont
  {Shibata}}, \bibinfo {author} {\bibfnamefont {K.}~\bibnamefont {Ueda}},
  \bibinfo {author} {\bibfnamefont {T.}~\bibnamefont {Nishino}}, \ and\
  \bibinfo {author} {\bibfnamefont {C.}~\bibnamefont {Ishii}},\ }\href@noop {}
  {\bibfield  {journal} {\bibinfo  {journal} {Physical Review B}\ }\textbf
  {\bibinfo {volume} {54}},\ \bibinfo {pages} {13495} (\bibinfo {year}
  {1996})}\BibitemShut {NoStop}%
\bibitem [{\citenamefont {Shibata}\ \emph {et~al.}(1997)\citenamefont
  {Shibata}, \citenamefont {Tsvelik},\ and\ \citenamefont
  {Ueda}}]{shibata1997one}%
  \BibitemOpen
  \bibfield  {author} {\bibinfo {author} {\bibfnamefont {N.}~\bibnamefont
  {Shibata}}, \bibinfo {author} {\bibfnamefont {A.}~\bibnamefont {Tsvelik}}, \
  and\ \bibinfo {author} {\bibfnamefont {K.}~\bibnamefont {Ueda}},\ }\href@noop
  {} {\bibfield  {journal} {\bibinfo  {journal} {Physical Review B}\ }\textbf
  {\bibinfo {volume} {56}},\ \bibinfo {pages} {330} (\bibinfo {year}
  {1997})}\BibitemShut {NoStop}%
\bibitem [{\citenamefont {Xavier}\ \emph {et~al.}(2003)\citenamefont {Xavier},
  \citenamefont {Pereira}, \citenamefont {Miranda},\ and\ \citenamefont
  {Affleck}}]{xavier2003dimerization}%
  \BibitemOpen
  \bibfield  {author} {\bibinfo {author} {\bibfnamefont {J.~C.}\ \bibnamefont
  {Xavier}}, \bibinfo {author} {\bibfnamefont {R.~G.}\ \bibnamefont {Pereira}},
  \bibinfo {author} {\bibfnamefont {E.}~\bibnamefont {Miranda}}, \ and\
  \bibinfo {author} {\bibfnamefont {I.}~\bibnamefont {Affleck}},\ }\href@noop
  {} {\bibfield  {journal} {\bibinfo  {journal} {Physical review letters}\
  }\textbf {\bibinfo {volume} {90}},\ \bibinfo {pages} {247204} (\bibinfo
  {year} {2003})}\BibitemShut {NoStop}%
\bibitem [{\citenamefont {Huang}\ \emph {et~al.}(2019)\citenamefont {Huang},
  \citenamefont {Sheng},\ and\ \citenamefont {Ting}}]{huang2019charge}%
  \BibitemOpen
  \bibfield  {author} {\bibinfo {author} {\bibfnamefont {Y.}~\bibnamefont
  {Huang}}, \bibinfo {author} {\bibfnamefont {D.}~\bibnamefont {Sheng}}, \ and\
  \bibinfo {author} {\bibfnamefont {C.-S.}\ \bibnamefont {Ting}},\ }\href@noop
  {} {\bibfield  {journal} {\bibinfo  {journal} {Physical Review B}\ }\textbf
  {\bibinfo {volume} {99}},\ \bibinfo {pages} {195109} (\bibinfo {year}
  {2019})}\BibitemShut {NoStop}%
\bibitem [{\citenamefont {Tsunetsugu}\ \emph {et~al.}(1992)\citenamefont
  {Tsunetsugu}, \citenamefont {Hatsugai}, \citenamefont {Ueda},\ and\
  \citenamefont {Sigrist}}]{tsunetsugu1992spin}%
  \BibitemOpen
  \bibfield  {author} {\bibinfo {author} {\bibfnamefont {H.}~\bibnamefont
  {Tsunetsugu}}, \bibinfo {author} {\bibfnamefont {Y.}~\bibnamefont
  {Hatsugai}}, \bibinfo {author} {\bibfnamefont {K.}~\bibnamefont {Ueda}}, \
  and\ \bibinfo {author} {\bibfnamefont {M.}~\bibnamefont {Sigrist}},\
  }\href@noop {} {\bibfield  {journal} {\bibinfo  {journal} {Physical Review
  B}\ }\textbf {\bibinfo {volume} {46}},\ \bibinfo {pages} {3175} (\bibinfo
  {year} {1992})}\BibitemShut {NoStop}%
\bibitem [{\citenamefont {Tsvelik}\ and\ \citenamefont
  {Yevtushenko}(2019)}]{tsvelik2019physics}%
  \BibitemOpen
  \bibfield  {author} {\bibinfo {author} {\bibfnamefont {A.~M.}\ \bibnamefont
  {Tsvelik}}\ and\ \bibinfo {author} {\bibfnamefont {O.}~\bibnamefont
  {Yevtushenko}},\ }\href@noop {} {\bibfield  {journal} {\bibinfo  {journal}
  {Physical Review B}\ }\textbf {\bibinfo {volume} {100}},\ \bibinfo {pages}
  {165110} (\bibinfo {year} {2019})}\BibitemShut {NoStop}%
\bibitem [{\citenamefont {Xavier}\ and\ \citenamefont
  {Miranda}(2008)}]{PhysRevB.78.144406}%
  \BibitemOpen
  \bibfield  {author} {\bibinfo {author} {\bibfnamefont {J.~C.}\ \bibnamefont
  {Xavier}}\ and\ \bibinfo {author} {\bibfnamefont {E.}~\bibnamefont
  {Miranda}},\ }\href {\doibase 10.1103/PhysRevB.78.144406} {\bibfield
  {journal} {\bibinfo  {journal} {Phys. Rev. B}\ }\textbf {\bibinfo {volume}
  {78}},\ \bibinfo {pages} {144406} (\bibinfo {year} {2008})}\BibitemShut
  {NoStop}%
\bibitem [{\citenamefont {White}(1992)}]{PhysRevLett.69.2863}%
  \BibitemOpen
  \bibfield  {author} {\bibinfo {author} {\bibfnamefont {S.~R.}\ \bibnamefont
  {White}},\ }\href {\doibase 10.1103/PhysRevLett.69.2863} {\bibfield
  {journal} {\bibinfo  {journal} {Phys. Rev. Lett.}\ }\textbf {\bibinfo
  {volume} {69}},\ \bibinfo {pages} {2863} (\bibinfo {year}
  {1992})}\BibitemShut {NoStop}%
\bibitem [{\citenamefont {White}(1993)}]{PhysRevB.48.10345}%
  \BibitemOpen
  \bibfield  {author} {\bibinfo {author} {\bibfnamefont {S.~R.}\ \bibnamefont
  {White}},\ }\href {\doibase 10.1103/PhysRevB.48.10345} {\bibfield  {journal}
  {\bibinfo  {journal} {Phys. Rev. B}\ }\textbf {\bibinfo {volume} {48}},\
  \bibinfo {pages} {10345} (\bibinfo {year} {1993})}\BibitemShut {NoStop}%
\bibitem [{\citenamefont {Schollw{\"o}ck}(2011)}]{schollwock2011density}%
  \BibitemOpen
  \bibfield  {author} {\bibinfo {author} {\bibfnamefont {U.}~\bibnamefont
  {Schollw{\"o}ck}},\ }\href@noop {} {\bibfield  {journal} {\bibinfo  {journal}
  {Annals of Physics}\ }\textbf {\bibinfo {volume} {326}},\ \bibinfo {pages}
  {96} (\bibinfo {year} {2011})}\BibitemShut {NoStop}%
\bibitem [{\citenamefont {Haldane}(1982)}]{haldane1982spontaneous}%
  \BibitemOpen
  \bibfield  {author} {\bibinfo {author} {\bibfnamefont {F.}~\bibnamefont
  {Haldane}},\ }\href@noop {} {\bibfield  {journal} {\bibinfo  {journal}
  {Physical Review B}\ }\textbf {\bibinfo {volume} {25}},\ \bibinfo {pages}
  {4925} (\bibinfo {year} {1982})}\BibitemShut {NoStop}%
\bibitem [{ITe()}]{ITensorandTenPy}%
  \BibitemOpen
  \href@noop {} {}\bibinfo {howpublished} {Calculations were performed using
  the ITensor Library \url{http://itensor.org/} and the TeNPy library (version
  0.4.1)}\BibitemShut {NoStop}%
\bibitem [{\citenamefont {Hauschild}\ and\ \citenamefont
  {Pollmann}(2018)}]{tenpy}%
  \BibitemOpen
  \bibfield  {author} {\bibinfo {author} {\bibfnamefont {J.}~\bibnamefont
  {Hauschild}}\ and\ \bibinfo {author} {\bibfnamefont {F.}~\bibnamefont
  {Pollmann}},\ }\href {\doibase 10.21468/SciPostPhysLectNotes.5} {\bibfield
  {journal} {\bibinfo  {journal} {SciPost Phys. Lect. Notes}\ ,\ \bibinfo
  {pages} {5}} (\bibinfo {year} {2018})},\ \bibinfo {note} {code available from
  \url{https://github.com/tenpy/tenpy}},\ \Eprint
  {http://arxiv.org/abs/1805.00055} {arXiv:1805.00055} \BibitemShut {NoStop}%
\bibitem [{\citenamefont {White}\ and\ \citenamefont
  {Affleck}(1996)}]{white1996dimerization}%
  \BibitemOpen
  \bibfield  {author} {\bibinfo {author} {\bibfnamefont {S.~R.}\ \bibnamefont
  {White}}\ and\ \bibinfo {author} {\bibfnamefont {I.}~\bibnamefont
  {Affleck}},\ }\href@noop {} {\bibfield  {journal} {\bibinfo  {journal}
  {Physical Review B}\ }\textbf {\bibinfo {volume} {54}},\ \bibinfo {pages}
  {9862} (\bibinfo {year} {1996})}\BibitemShut {NoStop}%
\bibitem [{\citenamefont {Calabrese}\ and\ \citenamefont
  {Cardy}(2004)}]{calabrese2004entanglement}%
  \BibitemOpen
  \bibfield  {author} {\bibinfo {author} {\bibfnamefont {P.}~\bibnamefont
  {Calabrese}}\ and\ \bibinfo {author} {\bibfnamefont {J.}~\bibnamefont
  {Cardy}},\ }\href@noop {} {\bibfield  {journal} {\bibinfo  {journal} {Journal
  of Statistical Mechanics: Theory and Experiment}\ }\textbf {\bibinfo {volume}
  {2004}},\ \bibinfo {pages} {P06002} (\bibinfo {year} {2004})}\BibitemShut
  {NoStop}%
\bibitem [{\citenamefont {Ruderman}\ and\ \citenamefont
  {Kittel}(1954)}]{ruderman1954indirect}%
  \BibitemOpen
  \bibfield  {author} {\bibinfo {author} {\bibfnamefont {M.~A.}\ \bibnamefont
  {Ruderman}}\ and\ \bibinfo {author} {\bibfnamefont {C.}~\bibnamefont
  {Kittel}},\ }\href@noop {} {\bibfield  {journal} {\bibinfo  {journal}
  {Physical Review}\ }\textbf {\bibinfo {volume} {96}},\ \bibinfo {pages} {99}
  (\bibinfo {year} {1954})}\BibitemShut {NoStop}%
\bibitem [{\citenamefont {Kasuya}(1956)}]{kasuya1956theory}%
  \BibitemOpen
  \bibfield  {author} {\bibinfo {author} {\bibfnamefont {T.}~\bibnamefont
  {Kasuya}},\ }\href@noop {} {\bibfield  {journal} {\bibinfo  {journal}
  {Progress of theoretical physics}\ }\textbf {\bibinfo {volume} {16}},\
  \bibinfo {pages} {45} (\bibinfo {year} {1956})}\BibitemShut {NoStop}%
\bibitem [{\citenamefont {Yosida}(1957)}]{yosida1957magnetic}%
  \BibitemOpen
  \bibfield  {author} {\bibinfo {author} {\bibfnamefont {K.}~\bibnamefont
  {Yosida}},\ }\href@noop {} {\bibfield  {journal} {\bibinfo  {journal}
  {Physical Review}\ }\textbf {\bibinfo {volume} {106}},\ \bibinfo {pages}
  {893} (\bibinfo {year} {1957})}\BibitemShut {NoStop}%
\bibitem [{\citenamefont {Shibata}\ and\ \citenamefont
  {Hotta}(2011)}]{shibata2011boundary}%
  \BibitemOpen
  \bibfield  {author} {\bibinfo {author} {\bibfnamefont {N.}~\bibnamefont
  {Shibata}}\ and\ \bibinfo {author} {\bibfnamefont {C.}~\bibnamefont
  {Hotta}},\ }\href@noop {} {\bibfield  {journal} {\bibinfo  {journal}
  {Physical Review B}\ }\textbf {\bibinfo {volume} {84}},\ \bibinfo {pages}
  {115116} (\bibinfo {year} {2011})}\BibitemShut {NoStop}%
\bibitem [{\citenamefont {Mermin}\ and\ \citenamefont
  {Wagner}(1966)}]{mermin1966absence}%
  \BibitemOpen
  \bibfield  {author} {\bibinfo {author} {\bibfnamefont {N.~D.}\ \bibnamefont
  {Mermin}}\ and\ \bibinfo {author} {\bibfnamefont {H.}~\bibnamefont
  {Wagner}},\ }\href@noop {} {\bibfield  {journal} {\bibinfo  {journal}
  {Physical Review Letters}\ }\textbf {\bibinfo {volume} {17}},\ \bibinfo
  {pages} {1133} (\bibinfo {year} {1966})}\BibitemShut {NoStop}%
\bibitem [{\citenamefont {Litvinov}\ and\ \citenamefont
  {Dugaev}(1998)}]{litvinov1998rkky}%
  \BibitemOpen
  \bibfield  {author} {\bibinfo {author} {\bibfnamefont {V.}~\bibnamefont
  {Litvinov}}\ and\ \bibinfo {author} {\bibfnamefont {V.}~\bibnamefont
  {Dugaev}},\ }\href@noop {} {\bibfield  {journal} {\bibinfo  {journal}
  {Physical Review B}\ }\textbf {\bibinfo {volume} {58}},\ \bibinfo {pages}
  {3584} (\bibinfo {year} {1998})}\BibitemShut {NoStop}%
\bibitem [{\citenamefont {Rusin}\ and\ \citenamefont
  {Zawadzki}(2017)}]{rusin2017calculation}%
  \BibitemOpen
  \bibfield  {author} {\bibinfo {author} {\bibfnamefont {T.~M.}\ \bibnamefont
  {Rusin}}\ and\ \bibinfo {author} {\bibfnamefont {W.}~\bibnamefont
  {Zawadzki}},\ }\href@noop {} {\bibfield  {journal} {\bibinfo  {journal}
  {Journal of Magnetism and Magnetic Materials}\ }\textbf {\bibinfo {volume}
  {441}},\ \bibinfo {pages} {387} (\bibinfo {year} {2017})}\BibitemShut
  {NoStop}%
\bibitem [{\citenamefont {Hotta}\ and\ \citenamefont
  {Shibata}(2006)}]{hotta2006absence}%
  \BibitemOpen
  \bibfield  {author} {\bibinfo {author} {\bibfnamefont {C.}~\bibnamefont
  {Hotta}}\ and\ \bibinfo {author} {\bibfnamefont {N.}~\bibnamefont
  {Shibata}},\ }\href@noop {} {\bibfield  {journal} {\bibinfo  {journal}
  {Physica B: Condensed Matter}\ }\textbf {\bibinfo {volume} {378}},\ \bibinfo
  {pages} {1039} (\bibinfo {year} {2006})}\BibitemShut {NoStop}%
\bibitem [{\citenamefont {Giamarchi}(2003)}]{giamarchi2003quantum}%
  \BibitemOpen
  \bibfield  {author} {\bibinfo {author} {\bibfnamefont {T.}~\bibnamefont
  {Giamarchi}},\ }\href@noop {} {\emph {\bibinfo {title} {Quantum physics in
  one dimension}}},\ Vol.\ \bibinfo {volume} {121}\ (\bibinfo  {publisher}
  {Clarendon press},\ \bibinfo {year} {2003})\BibitemShut {NoStop}%
\bibitem [{\citenamefont {Graf}\ \emph
  {et~al.}(2004{\natexlab{a}})\citenamefont {Graf}, \citenamefont {Brooks},
  \citenamefont {Choi}, \citenamefont {Uji}, \citenamefont {Dias},
  \citenamefont {Almeida},\ and\ \citenamefont {Matos}}]{graf2004suppression}%
  \BibitemOpen
  \bibfield  {author} {\bibinfo {author} {\bibfnamefont {D.}~\bibnamefont
  {Graf}}, \bibinfo {author} {\bibfnamefont {J.}~\bibnamefont {Brooks}},
  \bibinfo {author} {\bibfnamefont {E.}~\bibnamefont {Choi}}, \bibinfo {author}
  {\bibfnamefont {S.}~\bibnamefont {Uji}}, \bibinfo {author} {\bibfnamefont
  {J.}~\bibnamefont {Dias}}, \bibinfo {author} {\bibfnamefont {M.}~\bibnamefont
  {Almeida}}, \ and\ \bibinfo {author} {\bibfnamefont {M.}~\bibnamefont
  {Matos}},\ }\href@noop {} {\bibfield  {journal} {\bibinfo  {journal}
  {Physical Review B}\ }\textbf {\bibinfo {volume} {69}},\ \bibinfo {pages}
  {125113} (\bibinfo {year} {2004}{\natexlab{a}})}\BibitemShut {NoStop}%
\bibitem [{\citenamefont {Graf}\ \emph
  {et~al.}(2004{\natexlab{b}})\citenamefont {Graf}, \citenamefont {Choi},
  \citenamefont {Brooks}, \citenamefont {Matos}, \citenamefont {Henriques},\
  and\ \citenamefont {Almeida}}]{graf2004high}%
  \BibitemOpen
  \bibfield  {author} {\bibinfo {author} {\bibfnamefont {D.}~\bibnamefont
  {Graf}}, \bibinfo {author} {\bibfnamefont {E.}~\bibnamefont {Choi}}, \bibinfo
  {author} {\bibfnamefont {J.}~\bibnamefont {Brooks}}, \bibinfo {author}
  {\bibfnamefont {M.}~\bibnamefont {Matos}}, \bibinfo {author} {\bibfnamefont
  {R.}~\bibnamefont {Henriques}}, \ and\ \bibinfo {author} {\bibfnamefont
  {M.}~\bibnamefont {Almeida}},\ }\href@noop {} {\bibfield  {journal} {\bibinfo
   {journal} {Physical review letters}\ }\textbf {\bibinfo {volume} {93}},\
  \bibinfo {pages} {076406} (\bibinfo {year} {2004}{\natexlab{b}})}\BibitemShut
  {NoStop}%
\bibitem [{\citenamefont {Gama}\ \emph {et~al.}(1993)\citenamefont {Gama},
  \citenamefont {Henriques}, \citenamefont {Bonfait}, \citenamefont {Almeida},
  \citenamefont {Ravy}, \citenamefont {Pouget},\ and\ \citenamefont
  {Alc{\'a}cer}}]{gama1993interplay}%
  \BibitemOpen
  \bibfield  {author} {\bibinfo {author} {\bibfnamefont {V.}~\bibnamefont
  {Gama}}, \bibinfo {author} {\bibfnamefont {R.}~\bibnamefont {Henriques}},
  \bibinfo {author} {\bibfnamefont {G.}~\bibnamefont {Bonfait}}, \bibinfo
  {author} {\bibfnamefont {M.}~\bibnamefont {Almeida}}, \bibinfo {author}
  {\bibfnamefont {S.}~\bibnamefont {Ravy}}, \bibinfo {author} {\bibfnamefont
  {J.}~\bibnamefont {Pouget}}, \ and\ \bibinfo {author} {\bibfnamefont
  {L.}~\bibnamefont {Alc{\'a}cer}},\ }\href@noop {} {\bibfield  {journal}
  {\bibinfo  {journal} {Molecular Crystals and Liquid Crystals Science and
  Technology. Section A. Molecular Crystals and Liquid Crystals}\ }\textbf
  {\bibinfo {volume} {234}},\ \bibinfo {pages} {171} (\bibinfo {year}
  {1993})}\BibitemShut {NoStop}%
\bibitem [{\citenamefont {Bonfait}\ \emph {et~al.}(1993)\citenamefont
  {Bonfait}, \citenamefont {Matos}, \citenamefont {Henriques},\ and\
  \citenamefont {Almeida}}]{bonfait1993spin}%
  \BibitemOpen
  \bibfield  {author} {\bibinfo {author} {\bibfnamefont {G.}~\bibnamefont
  {Bonfait}}, \bibinfo {author} {\bibfnamefont {M.}~\bibnamefont {Matos}},
  \bibinfo {author} {\bibfnamefont {R.}~\bibnamefont {Henriques}}, \ and\
  \bibinfo {author} {\bibfnamefont {M.}~\bibnamefont {Almeida}},\ }\href@noop
  {} {\bibfield  {journal} {\bibinfo  {journal} {Le Journal de Physique IV}\
  }\textbf {\bibinfo {volume} {3}},\ \bibinfo {pages} {C2} (\bibinfo {year}
  {1993})}\BibitemShut {NoStop}%
\end{thebibliography}%
